\documentclass[aps,prb,reprint,twocolumn,floatfix,superscriptaddress,amsmath,amssymb,amsfonts, nofootinbib]{revtex4-2}

\usepackage{url}
\usepackage{bm}
\usepackage{graphicx}
\usepackage{amsmath}
\usepackage{amstext}
\usepackage{amssymb}
\usepackage{amsfonts}
\usepackage{amsbsy}
\usepackage{verbatim}
\usepackage{color}
\usepackage[colorlinks=true, urlcolor=blue, linkcolor=blue, citecolor=blue, pdftex]{hyperref}
\usepackage{floatrow}
\usepackage{float}
\usepackage{tikz}
\usepackage{gensymb}
\usepackage{textcomp}
\usepackage{enumitem}
\usepackage[version=3]{mhchem}
\usepackage{afterpage}
\usepackage{pbox}
\usepackage{dsfont}
\usepackage{siunitx}
\usepackage{stackengine,wasysym,scalerel}
\usepackage{latexsym}
\usepackage{cancel}
\usepackage{comment}
\usepackage{array, multirow, bigdelim, makecell, booktabs}
\usepackage[misc]{ifsym}
\usepackage[capitalise]{cleveref}
\usepackage{times}

\usepackage{color}
\usepackage[colorlinks=true, urlcolor=blue, linkcolor=blue, citecolor=blue, pdftex]{hyperref}

\usepackage[normalem]{ulem}

\definecolor{darkblue}{HTML}{004D6B}
\definecolor{darkred}{HTML}{8c1515}
\definecolor{darkgreen}{HTML}{006400}
\usepackage{hyperref}
\hypersetup{
	colorlinks=true,
	urlcolor=darkblue,
	citecolor=darkblue,
	linkcolor=darkblue,
	breaklinks
}

\begin{document}
\title{Noncoplanar orders and quantum disordered states in maple-leaf (anti)ferromagnets}
 \author{Martin Gemb\'e}
  \altaffiliation{These authors contributed equally to this work.}
 \affiliation{Institute for Theoretical Physics, University of Cologne, 50937 Cologne, Germany}
 \author{Lasse Gresista}
  \altaffiliation{These authors contributed equally to this work.}
 \affiliation{Institute for Theoretical Physics, University of Cologne, 50937 Cologne, Germany}
 \affiliation{Department of Physics and Quantum Centre of Excellence for Diamond and Emergent Materials (QuCenDiEM), Indian Institute of Technology Madras, Chennai 600036, India}
 \author{Heinz-J\"{u}rgen Schmidt} 
 \affiliation{Fachbereich Mathematik, Informatik und Physik, Universit\"{a}t Osnabr\"{u}ck, 49069 Osnabr\"{u}ck, Germany}
 \author{Ciar\'an Hickey}
 \affiliation{Institute for Theoretical Physics, University of Cologne, 50937 Cologne, Germany}
 \affiliation{School of Physics, University College Dublin, Belfield, Dublin 4, Ireland}
 \affiliation{Centre for Quantum Engineering, Science, and Technology, University College Dublin, Dublin 4, Ireland}
 \author{Yasir Iqbal}
 \affiliation{Department of Physics and Quantum Centre of Excellence for Diamond and Emergent Materials (QuCenDiEM), Indian Institute of Technology Madras, Chennai 600036, India}
 \author{Simon Trebst}
 \affiliation{Institute for Theoretical Physics, University of Cologne, 50937 Cologne, Germany}

\begin{abstract} 
A promising route towards the realization of chiral spin liquids is the quantum melting of classically noncoplanar spin states via 
quantum fluctuations. 
In the classical realm, such noncoplanar orders can effectively be stabilized by interactions beyond nearest neighbors. 
Motivated by the recent synthesis of materials with a maple-leaf lattice geometry, we study the effect of cross-plaquette couplings on elementary Heisenberg antiferromagnets for this geometry (as well as their ferromagnetic counterparts). 
We find a rich spectrum of noncoplanar states, including a novel icosahedral order as well as incommensurate spin spirals,
using large-scale Monte Carlo simulations in combination with a semi-analytical analysis.
To inspect the potential quantum melting of these states, we analyze the quantum $S = 1/2$ variant of these models
using  pseudo-fermion functional renormalization group (pf-FRG) simulations.
Notably, we indeed find extended parameter regimes lacking long-range magnetic order -- in regions classically occupied by noncoplanar orders --
which we putatively identify with the possible formation of chiral quantum spin liquids. 
\end{abstract}

\date{\today}

\maketitle

\section{Introduction}



Magnetic interactions on geometrically frustrated lattices offer the possibility of realizing exotic spin textures such as topologically non-trivial skyrmion~\cite{Muhlbauer-2009,Nagaosa-2013} and hedgehog crystals~\cite{Binz-2006}, or regular magnetic orders with spins oriented towards the vertices of platonic solids~\cite{Messio-2011}. Owing to the noncoplanar character of the underlying spin configurations, these textures feature long-range order of scalar spin chirality $\mathbf{S}_1\cdot(\mathbf{S}_{2}\times\mathbf{S}_{3})$ defined by three localized spins~\cite{Oosterom-1983}. Interest in models with such ground states stems from the expectation that for small values of spin, such as $S=1/2$, the spin ordering could undergo quantum melting, while the long-range order in chirality persists in the resulting nonmagnetic ground state, giving rise to a chiral spin liquid~\cite{hickey2016,hickey2017,KalmeyerLaughlin1987,Savary-2017,bieri2016}. Traditionally, realization of such textures has required coupling with itinerant electrons~\cite{Martin-2008,Akagi-2010,Akagi-2012,Barros-2014,Hayami-2017,Ozawa-2017} or magnetic fields~\cite{Park-2011,Yang-2016,Shimokawa-2019,Mohylna-2022}, and invoked Dzyaloshinksii-Moriya~\cite{Bogdanov-1989,Yi-2009} or multi-spin~\cite{Momoi-1997,Kubo-1997,Cookmeyer-2021} interactions. Progressively, it has been shown that this plethora of noncoplanar magnetic orders can be stabilized in simple Heisenberg models with competing long-range interactions even in the absence of a magnetic field~\cite{Domenge-2005,Janson-2008,Domenge-2008,Messio-2011,Okubo-2012,Aoyama-2021,Aoyama-2022,Gembe2023}.

\begin{figure}[b]
	\centering
	\includegraphics[width=0.75\linewidth]{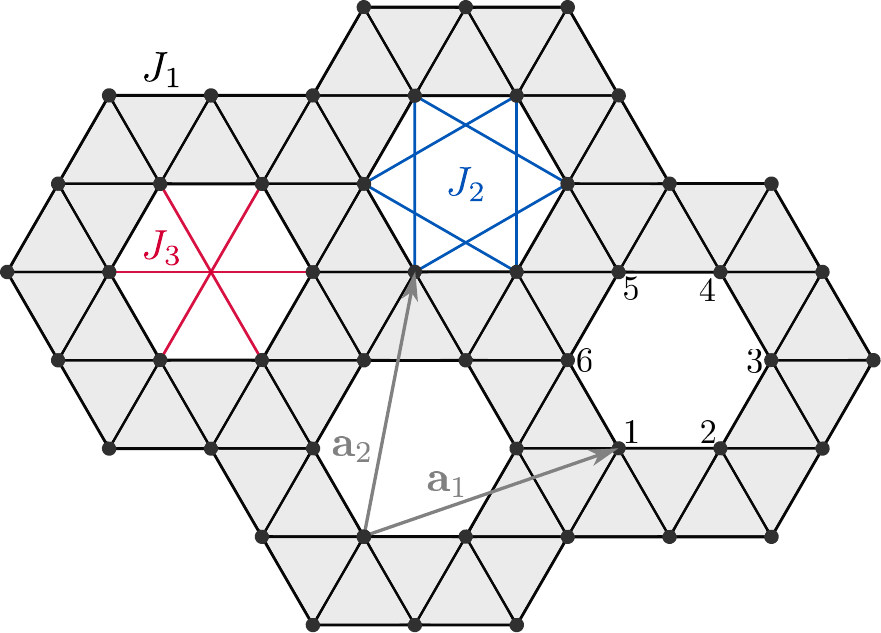}
	\caption{\textbf{Maple-leaf lattice} 
    	with nearest-neighbor coupling $J_1$ and further-neighbor cross-plaquette couplings $J_2$ and $J_3$. 
	The unit cell consists of the six sites that make up a hexagonal plaquette.
    	} 
	\label{Fig:maple_leaf_lattice}
\end{figure}

The exemplary textbook example of frustration is a triangular motif with antiferromagnetically interacting spins at its vertices~\cite{Balents2010}. Their edge-shared tessellation forms a triangular lattice which has been fertile ground for skyrmion physics~\cite{Okubo-2012,Karube-2016,Kurumaji-2019,Fang-2021,Mohylna-2022,Rosales-2015}. For a $S=1/2$ Heisenberg model with first $J_{1}$, second $J_{2}$ and third $J_{3}$ neighbor antiferromagnetic couplings on the triangular lattice there was an early proposal for a chiral spin liquid~\cite{Gong-2019}, which has lately been challenged~\cite{Jiang-2023}, and its existence remains debatable. Additional scalar spin chiral interactions need to be invoked to realize a stable chiral spin liquid phase which emerges out of quantum melting of noncoplanar tetrahedral order in the corresponding classical model~\cite{Gong-2017,Wietek-2017,hickey2017}. On the other hand, a corner-sharing tessellation of triangles, which leads to only a marginal alleviation of frustration, forms the kagome lattice \textemdash alternatively viewed as a $1/4$ site-depletion of the triangular lattice. Here, for antiferromagnetic $J_{1}$ and $J_{2}$, the inclusion of antiferromagnetic $J_{3}$ interactions across hexagons alone suffices to trigger a robust chiral spin liquid of the Kalmeyer-Laughlin type, possibly descending from quantum melting of parent cuboc orders~\cite{Hu-2015,Gong-2015,Wietek-2015,Fabrizio-2022,Bose-2023}, while for ferromagnetic $J_{1}$, a similar scenario has been argued for in Refs.~\cite{Bieri-2015,bieri2016,Iqbal-2015}. An analogous situation could potentially be realized on the square-kagome lattice where a multitude of noncoplanar orders, including novel cuboc states, have recently been reported in the classical Heisenberg model with competing long-range cross-plaquette interactions~\cite{Gembe2023}. Very recently, geometrical frustration inherent to various fullerene molecules has been shown to induce noncoplanar textures and chiral spin states in Heisenberg models~\cite{Szabo-2024}.

In this work, we consider a comparatively sparse $1/7$ site-depletion of the triangular lattice that leads to the five-fold coordinated maple-leaf lattice~\cite{Betts-1995,Misguich-1999,Schulenburg-2000,Schmalfuss-2002,Farnell-2011,Farnell-2014,Farnell-2018,ghosh2022,Gresista-2023,beck2024phase,ghosh2024spin,Ghosh_mag,schmoll2024} \textemdash\ a uniform tiling of triangles and hexagons, as shown in Fig.~\ref{Fig:maple_leaf_lattice} below. Hence, both in terms of the depletion density and coordination number, the maple-leaf lattice is intermediate between the triangular and kagome lattices, and one may wonder about the potential existence of a robust chiral spin liquid. As a guiding light in search of this phase on the maple-leaf lattice, it is important to first identify regions in parameter space of classical Heisenberg models which are host to non-coplanar magnetic orders. To this end, we consider a spatially isotropic Heisenberg model with both ferro- and antiferromagnetic $J_{1}$ and $J_{2}$ interactions, and, motivated by the kagome, include $J_{3}$ interactions across hexagonal plaquettes. Our study reveals a rich landscape comprised of four noncoplanar classical orders, together with two coplanar orders. In particular, this includes a previously unreported novel configuration where the spins point to the vertices of an icosahedron, which we dub {\it icosahedral} order, inspired by the cuboctohedral orders reported on the kagome lattice~\cite{Messio-2011}. Employing a state-of-the-art implementation of the pseudo-fermion functional renormalization group approach~\cite{muller2023}, we assess the impact of quantum fluctuations for $S=1/2$ and find an extended region in the $J_{2}$-$J_{3}$ plane which lacks long-range magnetic order. Importantly, the span of this nonmagnetic region encompasses regions classically occupied by noncoplanar orders, and could thus tentatively be associated with a putative chiral spin liquid. Furthermore, since the maple-leaf lattice lacks reflection symmetry about any straight line, the putative chiral spin liquid could lie outside the realm of the standard Kalmeyer-Laughlin paradigm which involves breaking of reflection symmetry up to time-reversal.  

This paper is structured as follows. We first introduce the model in Sec.~\ref{sec:model}. We then consider the antiferromagnetic case with $J_1 > 0$, for which the classical phase diagram is discussed in Sec.~\ref{sec:classical}, before we turn to the effects of  quantum fluctuations in Sec.~\ref{sec:quantum}. We conclude by discussing both the classical and quantum phase diagram for ferromagnetic $J_1 < 0$ in Sec.~\ref{sec:ferromagnet}.

\section{Model}
\label{sec:model}
The maple-leaf lattice \cite{Betts-1995} is an Archimedean lattice that is obtained by a periodic depletion of 1/7 of the sites of the triangular lattice, as visualized in Fig.~\ref{Fig:maple_leaf_lattice}. Its coordination number is $z=5$ and, therefore, it is intermediately frustrated between the kagome $(z=4)$ lattice and the triangular $(z=6)$ lattice. It can be described by the lattice vectors
\begin{align*}
    \mathbf{a}_1 &= \left(\frac{3\sqrt{3}}{2},-\frac{1}{2}\right), & \mathbf{a}_2 &= \left(\sqrt{3},2\right), 
\end{align*}
and a unit cell comprising six sites with relative coordinates 
\begin{align*}
    \mathbf{\delta}_1 &= \left(0,0\right), & \mathbf{\delta}_2 &= \left(\frac{\sqrt{3}}{2},-\frac{1}{2}\right), \nonumber \\ 
    \mathbf{\delta}_3 &= \left(\sqrt{3},0\right), & \mathbf{\delta}_4 &= \left(\sqrt{3},1\right),  \nonumber \\
    \mathbf{\delta}_5 &= \left(\frac{\sqrt{3}}{2},\frac{3}{2}\right), & \mathbf{\delta}_6 &= \left(0,1\right).
\end{align*}
With the three different couplings $J_1$ (nearest-neighbor), as well as $J_2$ and $J_3$ (cross-plaquette), as indicated in Fig.~\ref{Fig:maple_leaf_lattice}, the Heisenberg Hamiltonian on the maple-leaf lattice can be written as
\begin{align}
\label{eq:heisenberg_model}
  \mathcal{H} = \sum_{\langle{i,j}\rangle \in a} J_a\,  \mathbf{S}_i \cdot  \mathbf{S}_j \,,
\end{align}
where $a=\{1,2,3\}$ runs over the three different types of bonds.

\section{Classical Phase Diagram of the antiferromagnet}
\label{sec:classical}
To study the classical ground-state phase diagram of the Heisenberg model on the maple-leaf lattice, Eq.~\eqref{eq:heisenberg_model}, we employ large-scale classical Monte Carlo simulations, which we combine with a semi-analytical method that allows us to determine the exact phase boundaries (for details see Appendix~\ref{app:classical_phase_diagram}).

\begin{figure*}[t]
	\centering
	\includegraphics[width=0.9\linewidth]{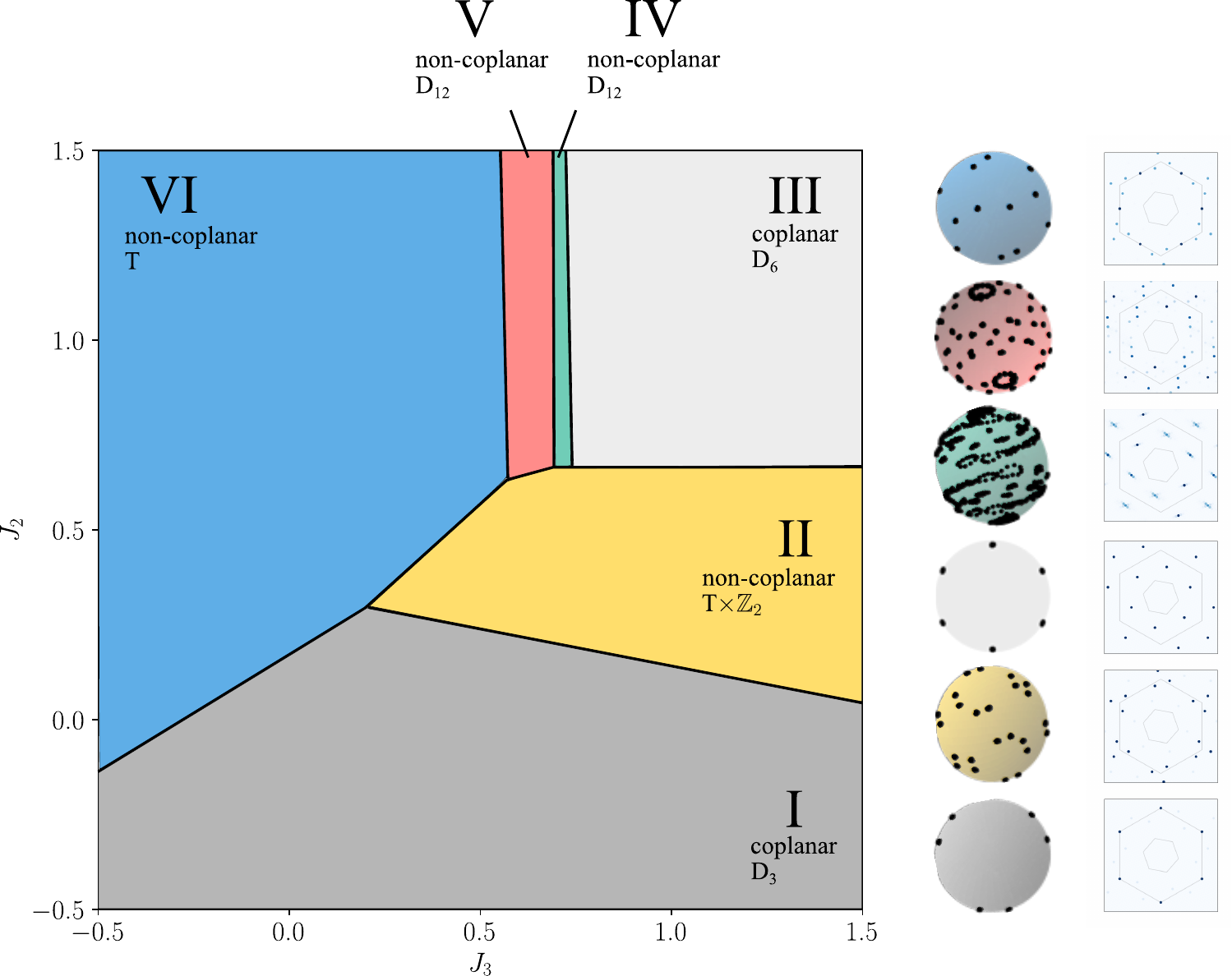}
	\caption{\textbf{Classical phase diagram of the antiferromagnet} 
    on the maple-leaf lattice with six different phases (labeled I-VI) as a function of $J_2$ and $J_3$ (and fixed nearest-neighbor coupling $J_1 = 1$). Besides indicating the phase boundaries (solid lines), the phases are described by 
	symmetry (with $D_n$ referring to the dihedral group of order $n$ and $T$ to tetrahedral symmetry), and coplanarity. Also, we show common origin plots and spin structure factors from Monte Carlo simulations ($T/J_1 = 10^{-4}, \ N = 864$) for each phase
	to the right of the phase diagram.}
	\label{Fig:classical_phase_diagram}
\end{figure*}

Upon varying the cross-plaquette interactions $J_2$ and $J_3$ (where $J_1 = 1$ is fixed to be antiferromagnetic), we indeed find a number of different ground-state phases, including coplanar and noncoplanar magnetic orders, as summarized in the classical phase diagram of Fig.~\ref{Fig:classical_phase_diagram}.  From the representative common origin plots of the ground-state real-space spin configurations (and also the static spin structure factors) next to the phase diagram, one can see at first glance that the six phases found (labeled I to VI) are clearly distinct from one another. 

The coplanar phases I and III appear in the form of two different six-sublattice ordered states, the first of which has already been described in the context of the quantum model on the maple-leaf lattice without cross-plaquette interactions. The remaining four phases come in different noncoplanar orders, including commensurate variants (phases II and VI) as well as incommensurate ones (phases IV and V). We will provide details about each phase in the remainder of this section. It is noteworthy that the ground states of all phases, with the exception of phases IV and V, can also be calculated analytically using the Luttinger-Tisza (LT) approach, see Appendix \ref{app:unconstrained-lt}. For phases IV and V, however, the LT approach yields unphysical ground states with a spin dimension greater than 3. 

We note that incommensurate spin configurations can not be faithfully captured with periodic boundary conditions on a finite lattice, as utilized by our Monte Carlo calculations and the subsequent semi-classical analysis. As will be discussed more thoroughly in Sec.~\ref{sec:quantum}, our pf-FRG analysis of the quantum model (which employs boundary conditions which effectively simulate the thermodynamic limit) does not observe two distinct phases IV and V. Instead, the structure factor evolves continuously from phase III to VI, showing different incommensurate ordering vectors in between. The precise spin configurations in phase IV and V may therefore be artefacts of the finite spin clusters necessitated by our Monte Carlo analysis. 

\subsection{Coplanar Orders}
We start with the coplanar ground states of phase I and phase III, which make up large parts of the lower half and the upper right corner of the classical phase diagram of Fig.~\ref{Fig:classical_phase_diagram} respectively.
\subsubsection{Coplanar phase I with $D_3$ symmetry}
This order has already been described as the ground state in the classical limit of the quantum maple-leaf Heisenberg model in Refs.~\cite{Schulenburg-2000,Schmalfuss-2002,Farnell-2011,Gresista-2023} in the special case $J_2 = J_3 = 0$ and is in full agreement with our numerical and analytical results. It consists of six sublattices of spins and a three times larger magnetic unit cell, as indicated in Fig.~\ref{Fig:mll_coplanar_orders}(a). Within one geometrical unit cell, neighboring spins form an angle 
\begin{equation}
    \alpha = \arctan\left(\frac{\sqrt{3}}{3+2J_3}\right) + \pi
\end{equation}
while next-nearest neighbors are parallel. Equivalent spins in two neighboring geometric unit cells are rotated by $2\pi/3$. Its energy can be given explicitly as    
\begin{equation}
    E_\text{I} = -\frac{1}{2} \left( 1 - 2J_2 +\sqrt{3+3J_3+J_3^2} \right) \,.
\end{equation}
The symmetry of this ground state is given by the dihedral group of order 3, $D_3$. For the special case $J_3 = -2$, the six different spin vectors form a regular hexagon with $D_6$ symmetry. In the case $J_3 = -1$, on the other hand, the ground state becomes a $120^\circ$ state which still has $D_3$ symmetry.

\subsubsection{Coplanar phase III with $D_6$ symmetry}
\begin{figure}[h]
	\centering
	\includegraphics[width=1.0\columnwidth]{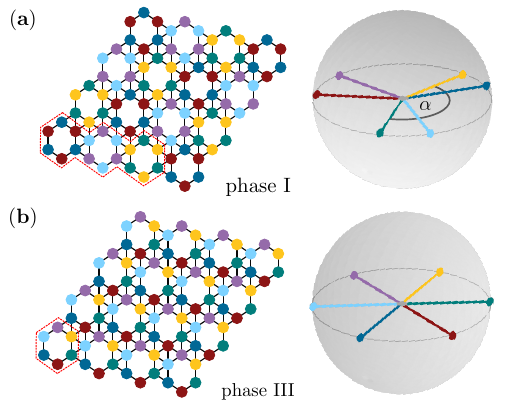}
	\caption{\textbf{Coplanar orders.} (a) The ground state of phase I has six sublattices of spins and a 18-site magnetic unit cell (outlined in red). Within one geometrical unit cell, neighboring spins form an angle $\alpha$ (for example, yellow and green spins) while next-nearest neighbors are parallel. Equivalent spins in two neighboring geometric unit cells are rotated by $2\pi/3$ (for example, yellow and light blue spins). (b) In phase III, the ground state has six sublattices as well, but the magnetic unit cell coincides with the geometrical unit cell. Within each unit cell, each spin points to a different corner of a regular hexagon and nearest neighbors (within a unit cell) form an angle of $\pi/3$.
    }	
	\label{Fig:mll_coplanar_orders}
\end{figure}

The rigid coplanar order of phase III, found in the upper right corner of the phase diagram, also has six sublattices of spins, but the magnetic unit cell coincides with the geometrical unit cell. Within every unit cell, each spin points to a different corner of a regular hexagon (see Fig.~\ref{Fig:mll_coplanar_orders}(b)) and nearest neighbors (within a unit cell) form an angle of $\pi/3$. The corresponding ground-state energy is
\begin{equation}
    E_\text{III} = -\frac{1}{2} \left( 1 + J_2 + J_3\right) \,.
\end{equation}
and the symmetry group of the ground state is the dihedral group of order 6, $D_6$.

\subsection{Noncoplanar Orders}
We now come to a discussion of the various noncoplanar magnetic orders, starting with the two commensurate structures found in phases II and VI of the phase diagram. The starting point for all semi-analytical descriptions are numerical Monte Carlo ground states at $T=10^{-4}$ with fixed $J_1 = 1$.
\subsubsection{Noncoplanar phase II with $T \times \mathbb{Z}_2$ symmetry}
The ground state of phase II consists of 24 sublattices of spins and the magnetic unit cell spans the same amount of sites, as shown exemplarily in Fig.~\ref{Fig:mll_non_coplanar_orders_1}(a)  for $J_2=1.0$ and $J_3 = 0.5$. A symmetry analysis of the ground state reveals that it is described by the symmetry group $T \times \mathbb{Z}_2$, the direct product of the tetrahedral group without reflections $T$ and the group $\lbrace \text{id}, -\text{id}\rbrace \cong \mathbb{Z}_2$. The energy of this state can be expressed as a function of two parameters $\lambda$ and $\alpha$ as
\begin{align}
    \notag E_\text{II} = \frac{1}{6} \Big[&-7+12\lambda^2-9J_2\lambda^2 +6J_2-3J_3 \\
    \notag & +2\lambda\sqrt{6-6\lambda^2}\cos\alpha+4\lambda^2\cos 2 \alpha  \\
    & + 2\lambda\sqrt{2-\lambda^2}\sin\alpha + 2\sqrt{3}\lambda^2\sin 2 \alpha\Big] \,. 
\end{align}
Minimizing this expression for $J_2=1.0$ and $J_3 = 0.5$ yields $E_{\text{II},\text{semi-analytical}} = -1.41782$ which fits well with the Monte Carlo result for the same parameters $E_{\text{II},\text{MC}} = -1.417(5)$. 
\subsubsection{Noncoplanar phase VI with $T$ symmetry}
\begin{figure}
	\centering
	\includegraphics[width=1.0\columnwidth]{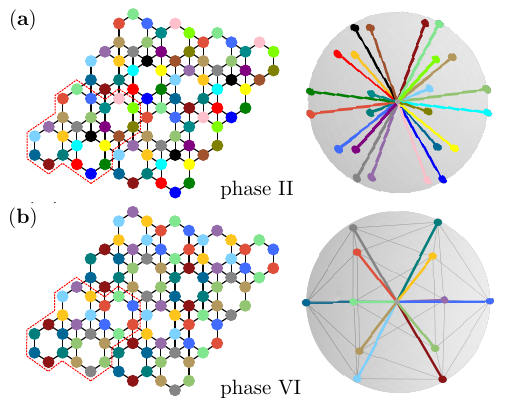}
	\caption{\textbf{Noncoplanar orders II and VI.} 
	(a) The ground state of phase II has 24 sublattices of spins and a magnetic unit cell with as many sites 
	(outlined in red in the real space arrangement on the left). Right: common origin plot of the 24 spin directions. 
	(b) The ground state of phase VI has a 24-site magnetic unit cell as well, but only 12 sublattices of spins. 
	In general, these point to the corners of a deformed icosahedron 
	which becomes regular in the special case $J_2 = 0.0, J_3 = -1.0$ (shown here).
	}
	\label{Fig:mll_non_coplanar_orders_1}
\end{figure}

The magnetic order of phase VI also has a 24-site magnetic unit cell, but only 12 sublattices of spins. In general, these point to the corners of a deformed \textit{icosahedron}, which becomes regular in the special case $J_2 = 0.0, J_3 = -1.0$ shown in Fig.~\ref{Fig:mll_non_coplanar_orders_1}(b). Its symmetry is described by the tetrahedral symmetry group without reflections $T$ (for the aforementioned special case it becomes the icosahedral symmetry group $I_h$). The general expression for the ground-state energy as a function of two parameters $\lambda$ and $\alpha$ is
\begin{align}
    \notag E_\text{VI} = \frac{1}{6} \Big[&3+6J_2-9J_2\lambda^2+3J_3+4\lambda\sqrt{6-6\lambda^2}\cos\alpha \\
    & +3\lambda^2\cos 2 \alpha -\sqrt{3}\lambda^2\sin2\alpha -6\lambda^2\Big] \,. 
\end{align}
Minimizing this term for, e.g., $J_2 = 0.5$ and $J_3 = -0.5$, leads to $E_{\text{VI},\text{semi-analytical}} = -1.60183$ in line with the corresponding Monte Carlo result $E_{\text{VI},\text{MC}} = -1.601(4)$. 

In the following, we conclude this report on the various noncoplanar orders on the maple-leaf lattice with the two incommensurate spiral structures of phases IV and V.

\begin{figure}[b]
	\centering
	\includegraphics[width=1\columnwidth]{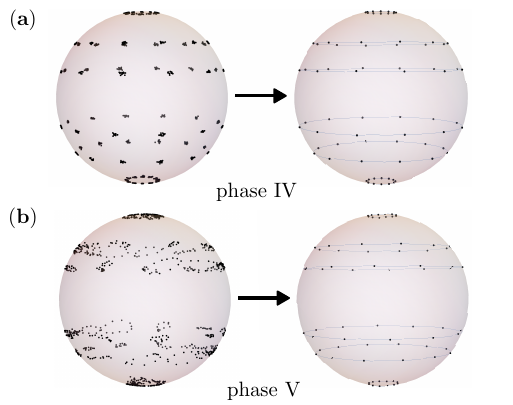}
	\caption{\textbf{Noncoplanar spiral orders IV and V.} (a) Left: common origin plot of the Monte Carlo ground state of spiral phase IV (for $J_2 = 1.0$ and $J_3 = 0.7$) with $N = 864$ spins. Right: These can be grouped into $M = 72$ unique directions and described analytically (see text). The resulting order has $D_{12}$ symmetry. (b) The same applies in the case of the spiral ground state of phase V, here shown for $J_2 = 1.0$ and $J_3 = 0.6$.
	}
	\label{Fig:mll_non_coplanar_orders_2}
\end{figure}

\subsubsection{Noncoplanar phase IV with $D_{12}$ symmetry}
In the spiral phase IV, as depicted in the common origin plot on the left of Fig.~\ref{Fig:mll_non_coplanar_orders_2}(a) for $J_2 = 1.0$ and $J_3 = 0.7$, we are left with $M = 72$ unique spin vectors after grouping those spins that point into the same direction (cf. the corresponding plot on the right of the very same figure). This state is symmetric under actions of the dihedral group of order 12, $D_{12}$, as well as under mirroring $z \mapsto -z$. The ground-state energy $E_\text{IV}$ can explicitly be expressed analytically as a function of eight parameters (which is too long to be specified here). Its minimization yields $E_{\text{IV},\text{semi-analytical}} = -1.35069$, consistent with the Monte Carlo result $E_{\text{IV},\text{MC}} = -1.350(5)$.  
\subsubsection{Noncoplanar phase V with $D_{12}$ symmetry}
For the spiral phase V, we consider the numerical Monte Carlo ground state for $J_2 = 1.0$ and $J_3 = 0.6$, as shown in the common origin plot on the left of Fig.~\ref{Fig:mll_non_coplanar_orders_2}(b). After grouping the $N = 864$ spin vectors according to their unique directions, we are left with $M = 72$ unique spin vectors, which can be divided into six groups with constant $z$-component, as visualized on the right of Fig.~\ref{Fig:mll_non_coplanar_orders_2}(b). Taking into account $D_{12}$ symmetry and invariance under $z \mapsto -z$, the state can be described by three $z$-components $z_1$, $z_2$, and $z_3$, and the difference angle $\delta\phi$ between the azimuth angles of the spins on the two lower circles on the one hand, and the upper circle on the other. With these parameters, the ground-state energy takes the form 
\begin{align}
    \notag E_\text{V} = \frac{1}{12} \Bigg[&-1 -4z_1^2 -z_2^2 - 2\sqrt{3-3z_2^2}\sqrt{1-z_3^2} \\ 
    \notag & - J_3\left( 1 + z_1^2 + 2z_2^2 + 2z_3^2 \right) \\
    \notag &- 4J_2 \Big( -z_2z_3 + z_1(z_2z_3)  \\
    \notag & + \sqrt{1-z_2^2}\sqrt{1-z_3^2}\Big) + \sqrt{1-z_1^2}\Bigg( \Big(2\sqrt{1-z_2^2}  \\
    \notag & - (2+\sqrt{3})J_2\big(\sqrt{1-z_2^2}-\sqrt{1-z_3^2}\big)\Big)\cos\delta\phi \\
    \notag & + \Big((-4+2\sqrt{3})\sqrt{1-z_2^2} \\
    & +J_2\big(\sqrt{1-z_2^2}-\sqrt{1-z_3^2}\big)\Big)\sin\delta\phi \Bigg)\Bigg]\,. 
\end{align}
Minimization of this energy yields $E_{\text{V},\text{semi-analytical}} = -1.32311$ in good agreement with and slightly smaller than the Monte Carlo result $E_{\text{V},\text{MC}} = -1.322(7)$ for the same parameters.

A compact summary that characterizes all six phases of the phase diagram of Fig.~\ref{Fig:classical_phase_diagram} by the ground-state symmetry, and corresponding
{\bf q} vectors, if any, is given in Table \ref{Tab:Summary}.

\begin{table}[h]
\begin{tabular}{c|c|c|ccc|c}
	\multirow{2}{*}{Phase} & \multirow{2}{*}{Symmetry} & \multirow{2}{*}{$\mathbf{q}$ vectors} & \multicolumn{4}{c}{Semi-analytical} \\
	  &   &   & $N$ & $M$ & $K$ & \# parms\\
	\hline \hline
	I 	& $D_3$	& $\left(\frac{2\pi}{3},- \frac{2\pi}{3}\right)$ & --& --& --&--\\
 	II 	& $T \times \mathbb{Z}_2$	& $(0,\pi),(\pi,0),(\pi,\pi)$ & --& --& --&--\\
  	III & $D_6$ & (0,0) & -- & -- & -- & --\\
   \hline
   	IV 	& $D_{12}$ & -- & 864 & 72 & 6 & 6\\
    V 	& $D_{12}$ & -- & 864 & 72 & 3 & 4\\
    \hline
    VI 	& $T$	& $(0,\pi),(\pi,0),(\pi,\pi)$  & --& --& --&--\\
\end{tabular}
\caption{{\bf Symmetry characterization of the six ground-state phases} of the phase diagram Fig.~\ref{Fig:classical_phase_diagram}.
		Given are the ground-state symmetry (second column), and the $\mathbf{q}$ vectors of each phase (if any, third column).
		 For the two semi-analytically described phases (IV and V), the compression of the parametrization of the spin spirals via clustering and symmetrization is given in the four columns on the right.
		 Technically, the semi-analytical description is obtained by starting with a common origin plot with $N=864$ spins sampled 
		 at $T=10^{-4}$ for a linear system size $L=12$, and is then given in terms of $M$, $K$, and the number of needed parameters (last column) to describe the phase (see Appendix \ref{app:classical_phase_diagram} for details).
}
\label{Tab:Summary}
\end{table}

\subsection{Thermodynamics}
\begin{figure}
	\centering
	\includegraphics[width=1\columnwidth]{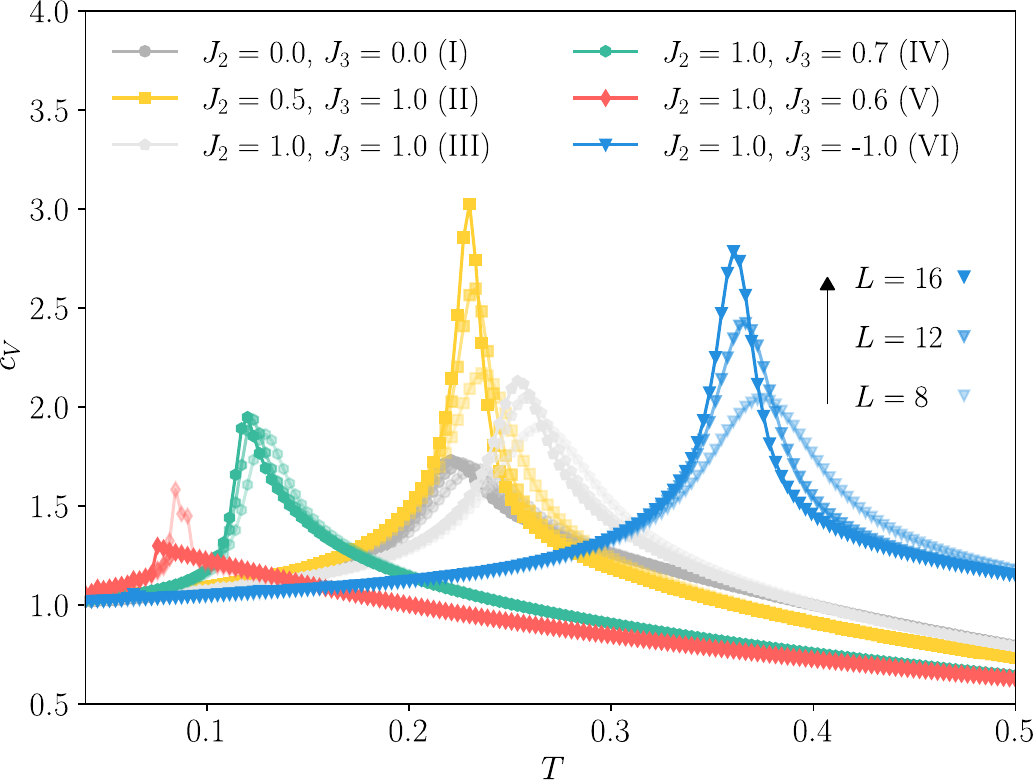}
	\caption{\textbf{Specific heat traces} from Monte Carlo simulations for all six phases, color-coded the same way as in the phase diagram Fig.~\ref{Fig:classical_phase_diagram}. The spiral phases IV and V (green and red, respectively) show complicated behavior indicating a sensitive dependence on finite-size effects. The four commensurate phases I (dark gray), II (yellow), III (light gray), and VI (blue), display a scaling behavior that is expected for a thermal phase transition with a single peak that grows with system size. Note, that for phase I, the simulated system sizes are $L = 9,12,15$ to ensure commensurability of the underlying order with the lattice.}
	\label{Fig:mll_td_all_phases}
\end{figure}

To round up the discussion of the classical phase diagram, specific heat traces for the six phases are shown in Fig.~\ref{Fig:mll_td_all_phases}. Peaks in the specific heat and their finite-size scaling can give us indications about the nature of the transition into the ordered ground state. Generally, a pronounced peak at a finite temperature $T_c$ whose height diverges with increasing system size is associated with a thermal phase transition into an ordered state. The Mermin-Wagner theorem generally forbids the associated breaking of continuous symmetries in two dimensional Heisenberg models. In the noncoplanar, i.e. chiral, phases, however, selecting a specific chirality breaks a discrete symmetry, and a true thermal phase transition may occur. The specific heat traces of phases II and VI indicate such a behavior showing a pronounced peak that grows significantly with system size. The specific heat in the coplanar phases I and III also shows a peak, but the height seems to saturate with increasing system size, indicating a thermal crossover. The spiral phases IV and V show complicated behavior; due to their incommensurability, these phases depend very sensitively on finite-size effects, and the nature of the phase transition can not be clearly determined, although their chiral nature would in principle allow for a finite temperature phase transition. 

\section{Quantum Phase Diagram of the antiferromagnet}
\label{sec:quantum}

\begin{figure*}
    \includegraphics{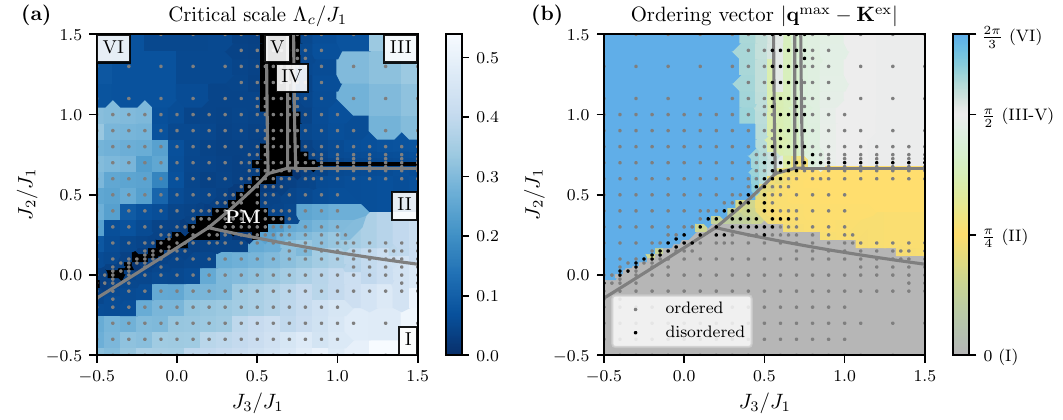}
    \includegraphics{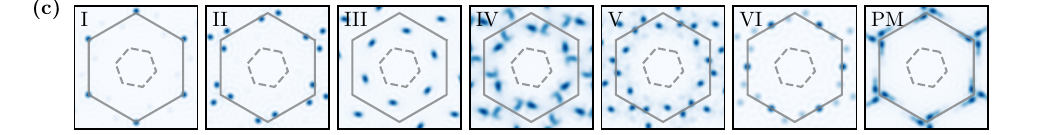}
    \caption{{\bf Quantum phase diagram for the antiferromagnet} as obtained from pf-FRG. \textbf{(a)} Critical scale $\Lambda_c$ where the renormalization group flow exhibits an instability, signaling the formation of long-range order. In the black regions the flow shows no instability ($\Lambda_c = 0$), indicating a disordered ground state. The gray dots indicate the couplings for which calculations were performed. \textbf{(b)} Distance of the momentum where the structure factor is maximal $\mathbf{q}^\mathrm{max}$ to the $\mathbf{K}^\mathrm{ex}$-point of the extended Brillouin zone. Quantum disordered points that show no flow breakdown $(\Lambda_c = 0)$ are colored black. The gray lines are the classical phase boundaries. \textbf{(c)} Structure factors at the same parameters as shown for the classical case in Fig.~\ref{Fig:classical_phase_diagram}. The dashed and solid gray lines show the first and  extended Brillouin zones, respectively.}
    \label{Fig:quantum-phase-diagram}
\end{figure*}

\begin{figure*}
    \centering
    \includegraphics{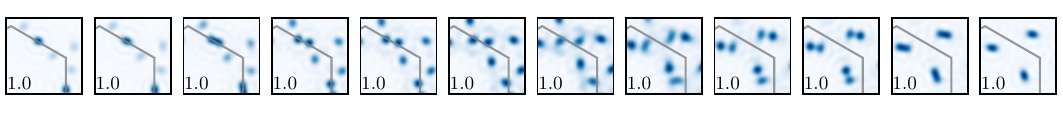}
    \caption{{\bf Quantum structure factor evolving from phase III to VI} for fixed $J_2/J_1 = 1$ and varying $J_3/J_1$ (indicated in the bottom left corners). We observe a seemingly continuous evolution with no clearly distinct signatures of phases V and VI.}
    \label{fig:quantum-sfs-transition}
\end{figure*}

In the previous section we unveiled the existence of a plethora of classically ordered phases in the maple-leaf model defined in Fig.~\ref{Fig:maple_leaf_lattice}. Most notably, we identified four phases (II, IV, V and VI) with a noncoplanar ground state. Now, we turn to the question of how these phases are affected by quantum fluctuations. Particular interest lies in finding parameter regimes where fluctuations melt the classical noncoplanar order into a ground state with restored spin rotational symmetry. Such a state would then be a strong candidate for a chiral quantum spin liquid \cite{bieri2016, hickey2016, hickey2017}. To achieve this, we replace the classical spins in the original model by $S=1/2$ spin operators. We then calculate the ground-state phase diagram of the resulting quantum model using the pseudo-fermion functional renormalization group (pf-FRG), a by now well-established method for distinguishing between magnetically ordered and disordered regimes at zero temperature \cite{muller2023}.

To probe for magnetic order in the ground state of a given spin Hamiltonian, we use the pf-FRG to calculate the $\emph{flow}$ of the (static) spin-spin correlation defined as\footnote{The spin-rotational symmetry of the Hamiltonian is preserved in the FRG flow. The $xx$, $yy$ and $zz$-correlations are therefore equivalent and it suffices to study just one of them.}
\begin{equation}
    \label{eq:correlations}
    \chi_{ij}^\Lambda = \int_0^\infty\!\!\! d\tau e^{i\omega \tau}\! \left\langle \hat{T}_\tau \hat{S}^z_i(\tau)\hat{S}^z_j(0)\right\rangle\Big|^\Lambda_{\omega = 0}
\end{equation}
where $\hat{T}_{\tau}$ is the time-ordering operator in imaginary time $\tau$ and $\Lambda$ is an infrared cutoff, or \emph{RG scale}, artificially introduced into the theory. A Fourier transformation then leads to the flow of the magnetic structure factor. If the ground state of the Hamiltonian under consideration exhibits magnetic order, this flow will exhibit a divergence, or \emph{flow breakdown}, at a finite \emph{critical scale} $\Lambda_c$ for the Bragg momenta characterizing the incipient order. Conversely, in the absence of a flow breakdown, the ground state is anticipated to be a disordered state, indicative of a potential quantum spin liquid.  More details on the pf-FRG and the criteria used to distinguish ordered from disordered states are provided in Appendix~\ref{app:pf-FRG}. 

In practice, we calculate the flow using the \texttt{PFFRGSolver.jl} Julia package, featuring state-of-the-art, adaptive integration schemes \cite{PFFRGSolver} for solving the pf-FRG flow equations. To discretize the continuous Matsubara frequency dependence of the four-point correlators, we choose an adaptive grid of $n_\Omega = 40$ bosonic and $n_\nu = 35 \times 35$ fermionic Matsubara frequencies. We use lattice truncations of up to $L = 15$, i.e., correlations are set to zero beyond a bond distance of 15. Using this setup, we calculate the quantum analog to the classical phase diagram in Fig.~\ref{Fig:classical_phase_diagram}, with a focus on identifying quantum disordered parameter regimes. The result for the critical scale is shown in Fig.~\ref{Fig:quantum-phase-diagram}(a), while Fig.~\ref{Fig:quantum-phase-diagram}(b) shows the ordering vectors (i.e. the momenta where the structure factor is maximal) and Fig.\ref{Fig:quantum-phase-diagram}(c) illustrates instances of the complete structure factor within the various phases we have identified.

\subsection{Quantum structure factors}
Before discussing the possibility of quantum disordered phases, let us first compare the indication of order visible in the pf-FRG structure factors with our classical analysis. Not surprisingly, the ground-state structure factors agree well with the classical case deep in the phases I, II, III and VI [compare Fig.~\ref{Fig:classical_phase_diagram} with Fig.~\ref{Fig:quantum-phase-diagram}(c)]. Nonetheless, in  proximity to certain phase boundaries, notable deviations emerge. 

Most prominently, within the region between phases III and VI, we don't observe two distinct phases IV and V, as identified in the classical model. Instead, as depicted in Fig.~\ref{fig:quantum-sfs-transition}, the structure factor continuously evolves from phase III to phase VI, showing peaks at incommensurate momenta in between. A similar situation arises in proximity to the triple point where phases I, II and VI converge. Here, the pf-FRG again reveals an extended region with a structure factor maximum at an incommensurate momentum. 

Incommensurate spin configurations can not be faithfully captured with periodic boundary conditions on a finite lattice, as utilized by our Monte Carlo calculations. This limitation is likely why, in the classical analysis, we identified only distinct phases with finite magnetic unit cells. In contrast, the pf-FRG employs open boundary conditions and thus avoids this issue. We note that, in both incommensurate regimes, the ordering vectors from pf-FRG align remarkably well with the momenta $\mathbf{q}^\mathrm{min}$ that minimize the energy in an unconstrained Luttinger-Tisza (LT) approach \cite{luttinger1946, luttinger1951}. In this LT approach, the constraint of constant spin length is softened enabling a straight-forward diagonalization of the Hamiltonian in momentum space (see Appendix~\ref{app:unconstrained-lt} for details). It has been argued that this approach provides an improved approximation to the quantum problem compared to a purely classical analysis \cite{kimchi2014}. More importantly, it takes into account the full infinite lattice, allowing for the description of both commensurate and incommensurate ground-state orders.

\subsection{Disordered phases}
Returning to the discussion of disordered, putative quantum spin liquid phases, they exactly seem to appear in the para\-meter regions where the classical and quantum structure factors disagree. In the pf-FRG, these phases manifest themselves by the absence of a flow breakdown ($\Lambda_c = 0$), illustrated by the black colored regions in Fig.~\ref{Fig:quantum-phase-diagram}(a). Most prominently, we observe an extended quantum disordered regime close to the triple point where the phases I, II and VI converge, which extends further into the classical noncoplanar phase II with $T \times \mathbb{Z}_2$ symmetry. Here, the interplay of quantum fluctuations and the competition of three neighboring phases seems to suppress the magnetic order, making the regime a promising candidate for an extended chiral quantum spin liquid phase. 

Furthermore, the incommensurate regime between phases III and VI also exhibits a vanishing critical scale. However, in this case the continuous evolution of the structure factor, and the resulting proximity to a spectrum of many different orders at any given point along the evolution, may pose a challenge for the pf-FRG to correctly identify a flow breakdown at a specific ordering vector. Incommensurate states also tend to have a flow breakdown at lower critical scales, further complicating the numerical identification. We can, therefore, not clearly determine whether this region is genuinely quantum disordered or if it is an artifact of our calculation.

\section{Phase diagram of the ferromagnet}
\label{sec:ferromagnet}

\begin{figure*}
    \centering
    \includegraphics[]{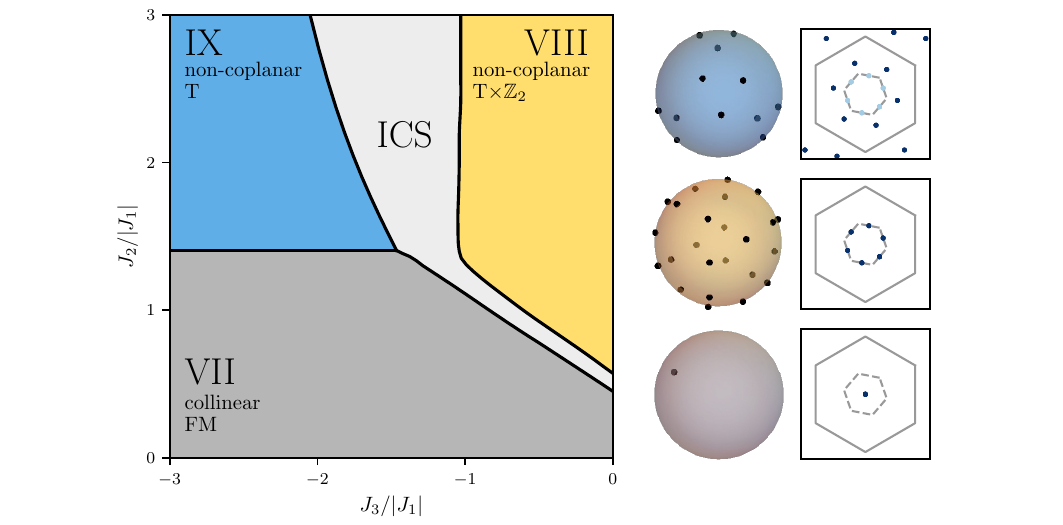}
    \caption{{\bf Classical phase diagram of the ferromagnet} as a function of $J_2$ and $J_3$ with fixed $J_1 < 0$. Phase VII, VIII, IX and the phase boundary between VII and IX are analytically calculated using the Luttinger-Tisza method. Phase VII exhibits ferromagnetic order, while phases VIII and IX are noncoplanar phases, whose common origin plot and classical structure factor are shown on the right. In the region labeled "ICS" the Luttinger-Tisza approach yields only unphysical states, where the hard spin constraint is not fulfilled. The corresponding $\mathbf{q}$-values occur at different incommensurate momenta that agree well with structure factor peaks from pf-FRG. This indicates a large incommensurate region likely hosting multiple different incommensurate ground states.}
    \label{fig:classical-phase-diagram-fm} 
\end{figure*}

\begin{figure}
    \centering
    \includegraphics{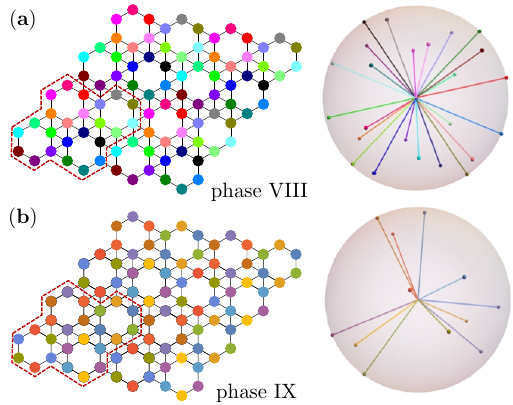}
    \caption{{\bf Noncoplanar orders of the ferromagnet} \textbf{(a)} The ground state of phase VIII has 24 sublattices of spins and a magnetic unit cell with as many sites (outlined in red in the real space arrangement on the left). The spin directions are symmetric under the symmetry group $T \times \mathbb{Z}$.
	\textbf{(b)} The ground state of phase IX has a 24-site magnetic unit cell, but only 12 sublattices of spins. These point to the corners of a deformed truncated tetrahedron and are symmetric under the symmetry group $T$.}
    \label{fig:fm-ncp-configurations}
\end{figure}

In the previous sections we identified multiple noncoplanar phases and potential chiral spin liquid candidates for the antiferromagnet with $J_1 > 0$. Motivated by these finding, we now examine the ferromagnetic case $J_1 < 0$. Focusing on regions of noncoplanar order, we first discuss the classical phase diagram, revealing two new noncoplanar phases, and subsequently consider the effects of quantum fluctuations.

\subsection{Classical phases}
We calculate the classical phase diagram using the Luttinger Tisza method. This allows us to capture three phases analytically: The ferromagnetic phase VII and the two noncoplanar phases VIII and IX. Their phase boundaries, structure factors and common-origin plots are shown in Fig.~\ref{fig:classical-phase-diagram-fm}. In the parameter region separating these phases (labeled "ICS"), the Luttinger-Tisza method shows $\mathbf{q}$-vectors at different incommensurate momenta, but only produces unphysical states that do not satisfy the hard spin constraint. Precisely characterizing the ground state orders in this incommensurate regime is challenging with classical Monte Carlo simulations due to its reliance on a finite periodic lattice. Since our focus is on identifying noncoplanar phases and potential chiral spin liquid candidates, we do not further resolve this regime using classical Monte Carlo. A figure illustrating how the $\mathbf{q}$-values from unconstrained Luttinger Tisza evolve is shown in appendix \ref{app:fm}, which align well with their quantum counterparts from pf-FRG, 
 discussed in the following section. Here, we first detail the classical phases we were able to capture analytically, before discussing the effect of quantum fluctuations in the next section.


\subsubsection*{Noncoplanar orders}
Both noncoplanar orders can be analytically calculated using the Luttinger-Tisza method and are characterized by  $\mathbf{q}$-vectors of $(0, \pi), (\pi, 0), (\pi, \pi)$ for the three spin components $S^x, S^y, S^z$, respectively. Their real-space spin configurations are illustrated in Fig.~\ref{fig:fm-ncp-configurations}. 

In phase VIII the magnetic unit cell contains 24 sites, with each spin pointing in one of 24 different directions symmetric under the group $T \times \mathbb{Z}_2$. 
The spin configuration is not rigid in the sense that it varies with changes in $J_2$ or $J_3$. It does, however, always stay $T \times \mathbb{Z}_2$ symmetric and retains the  qualitative form shown in Fig.~\ref{fig:classical-phase-diagram-fm}. The ground-state energy satisfies the equation (setting $J_1=-1$)
\begin{equation}
    \begin{aligned}
        -&1 - 8 J_2 - 9 J_2^2 - 2 J_2^3 - 9 J_3\\
        - &8 J_2 J_3 - 3 J_2^2 J_3 + J_3^2 + J_3^3\\ 
      + &(-9 - 8 J_2 - 3 J_2^2 + 2 J_3 + 3 J_3^2) 2E_\mathrm{VIII}\\
      + &(1 + 3 J_3) 4E_\mathrm{VIII}^2 + 8E_\mathrm{VIII}^3 = 0,
    \end{aligned}
\end{equation}
which evaluates to $E_\mathrm{VIII} = -2.97756$ for  $J_2/|J_1| = 3$ and $J_3/|J_1| = -\frac{1}{2}$. The  structure factor and real-space configuration for these parameters are shown in Figs.~\ref{fig:classical-phase-diagram-fm}~and~\ref{fig:fm-ncp-configurations}.

The order in phase IX also has a magnetic unit cell with 24 sites, but with only 12 different spin directions, which are permuted under the action of the tetrahedral group $T$. These spin directions form a polyhedron spanned by four equilateral triangles, which we call a “deformed truncated tetrahedron”.  The term “truncated tetrahedron” is known to refer to the Archimedean solid with 12 vertices and a surface consisting of four regular hexagons and four equilateral triangles. However, we can prove that the truncated tetrahedron is not realized in phase IX. The spin configuration within this phase remains constant when $J_3$ is varied, but the size and orientation of the equilateral triangles depends on the coupling $J_2$.
The ground-state energy satisfies the equation (setting $J_1=-1$)
\begin{equation}
\begin{aligned}
    &5 - 4 J_2 + 5 J_2^2 - 2 J_2^3\\
    + &5 J_3 - 4 J_2 J_3 + 3 J_2^2 J_3 - J_3^2 - J_3^3 \\
    + &(-5 + 4 J_2 - 3 J_2^2 + 2 J_3 + 3 J_3^2) 2E_\mathrm{IX}\\
    + &4(-1 - 3 J_3) E_\mathrm{IX}^2 + 8E_\mathrm{IX}^3 = 0.
\end{aligned}
\end{equation}
This, e.g., evaluates to $E_\mathrm{IX}=-3.22575$ for $J_2 = 3|J_1|$ and $J_3 = -3 |J_1|$ for which the structure factor and real-space configuration is shown in Figs.~\ref{fig:classical-phase-diagram-fm}~and~\ref{fig:fm-ncp-configurations}.
The boundary between phase IX and the ferromagnetic phase VII is given by the horizontal line 
$J_2/|J_1|= \frac{1}{9} \left(17-\sqrt{19}\right)\approx 1.40457$, see Fig.~\ref{fig:classical-phase-diagram-fm}.

\subsection{Effects of quantum fluctuations}
\label{sec:pffrg-fm}

\begin{figure*}
    \includegraphics{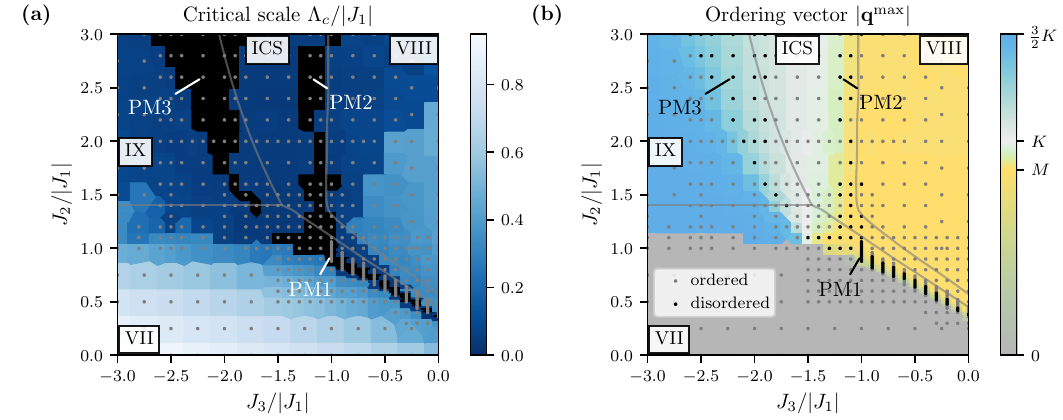}
    \includegraphics{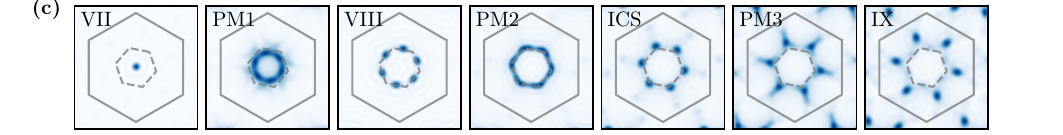}
    \caption{{\bf Quantum phase diagram for ferromagnetic $J_1 < 0$} as obtained from the pf-FRG. \textbf{(a)} Critical scale $\Lambda_c$ where the renormalization group flow exhibits an instability, signaling the formation of long-range order. In the black regions the flow shows no instability ($\Lambda_c = 0$), indicating a disordered ground state. The gray dots indicate the couplings for which calculations were performed. \textbf{(b)} Distance of the momentum where the structure factor is maximal $\mathbf{q}^\mathrm{max}$ to the $\mathbf{Gamma}$-point of the Brillouin zone. Quantum disordered points that show no flow breakdown $(\Lambda_c = 0)$ are colored black. The gray lines are the classical phase boundaries. \textbf{(c)} Structure factors deep in the three phases VII, VIII and IX, and in the incommensurate regions in between. The annotations indicate the parameters for which the structure factors are shown. The dashed and solid gray lines show the first and  extended Brillouin zones, respectively.}
    \label{Fig:quantum-phase-diagram-fm}
\end{figure*}

Employing pf-FRG simulations, we investigate the effects of quantum fluctuations on the phase diagram for the ferromagnetic case. Fig.~\ref{Fig:quantum-phase-diagram-fm} shows the resulting critical scale and ordering vectors, as well as structure factors for parameters inside the different observed phases. Compared to the classical case, we observe a relatively large shift of the phase boundaries: The ferromagnetic phase VII shrinks along all its phase boundaries. The incommensurate region separating phase VIII and IX enlarges and extends further into the classical phase IX. 

In the incommensurate regime, the structure factor seems to evolve continuously between the bordering commensurate phases. In the ICS region between phase VII and phase VIII, the structure factor peaks move continuously from the $\mathbf{\Gamma}$-point to the $\mathbf{M}$-point. Between phases VIII and IX the peaks move from the $\mathbf{M}$-point over the $\mathbf{K}$-point to the vector $\frac{3}{2}\mathbf{K}$. Note that both pf-FRG and unconstrained Luttinger-Tisza reveal no extended region where the $\mathbf{q}$-vector lies exactly at the $\mathbf{K}$-point, which would indicate commensurate order. We provide additional figures showing the evolution of structure factors in the incommensurate regions and cuts through the phase diagram in appendix~\ref{app:fm}.

In the incommensurate regime, we find three paramagnetic phases that exhibit no flow-breakdown in the FRG analysis, indicating a disordered ground state. All of the paramagnetic phases are situated at the border of, or even extend into one of the noncoplanar phases, making them candidates for realizing chiral quantum spin liquid ground states. Notably, the phases PM1 and PM2 display broad features rather than sharp peaks in the structure factor, which may be an additional indicator of quantum spin liquid behavior.

\section{Conclusions and outlook}
\label{sec:discussion}

We explore the classical and quantum phase diagram of the spatially isotropic Heisenberg ferro- and antiferromagnet on the maple-leaf lattice in the presence of long-range interactions. In search of noncoplanar magnetic orders, we show that the minimal set of couplings that need to be invoked to stabilize these orders are cross-hexagonal third neighbor interactions on top of the nearest-neighbor Heisenberg model -- similar to the kagome lattice where such couplings are known to trigger cuboc orders. A comprehensive classical Monte Carlo study finds a rich variety of noncoplanar states, including a new type of order wherein the spins point to the vertices of an icosahedron (dubbed {\it icosahedral} order), as well as complex incommensurate noncoplanar spirals. These states feature large magnetic unit cells with a highly intricate structure, but with the salient feature that they lend themselves to a semi\-ana\-lyti\-cal construction. We provide an optimal parameterization of the spin configuration of these states based on a careful symmetry analysis. This allows for obtaining explicit expressions (depending on only a few parameters) for their ground-state energy as a function of the interactions, which in turn permits us to accurately establish the phase boundaries between these complex phases. It is highly plausible that considering a more generalized symmetry-allowed model with all three couplings at first, second, and third neighbors being different with possible ferro- and antiferromagnetic combinations, would give us access to a comparatively richer landscape of exotic noncoplanar orders.

In addition to the ground state, we study the thermodynamics of noncoplanar states using classical Monte Carlo simulations. The behavior of the specific heat points to a finite-temperature phase transition in this classical two-dimensional model, as expected due to the chiral nature of the noncoplanar ground states~\cite{Messio-2011}. In the limit of low spin values, e.g., $S=1/2$, one may expect that strong quantum fluctuations preclude the formation of long-range magnetic order while the long-range order in chirality survives, thus possibly stabilizing a chiral spin liquid. Given that the maple-leaf lattice lacks reflection symmetry about any straight line, the issue concerning which lattice symmetries could be broken up to time-reversal (i.e., allowed chiral classes) in order to respect ``$PT$'' theorem (for $U(1)$ QSLs) is worth examining. Identifying the allowed symmetry patterns and microscopic nature of this putative QSL phase would involve a systematic projective symmetry group classification of chiral mean-field Ans\"atze with $U(1)$ and $\mathbb{Z}_{2}$ low-energy gauge groups~\cite{bieri2016,Sonnenschein-2024}. The ground-state energies and correlation functions of the corresponding Gutzwiller projected states could then be obtained within a variational Monte Carlo scheme~\cite{Becca_Sorella_2017,Ferrari-2023}, and compared to the structure factors obtained from pf-FRG in the current work. Alternatively, these Ans\"atze could be analyzed within a pf-FRG framework itself by performing a self-consistent Fock-like renormalized mean-field scheme to compute low-energy theories for emergent spinon excitations but using effective vertex functions instead of the bare couplings~\cite{Hering-2019}. Indeed, (gapless) $U(1)$ chiral spin liquids displaying cuboc type magnetic correlation profiles have been reported to be energetically competitive variational ground states in the $S=1/2$ $J_{1}$-$J_{2}$-$J_{d}$ Heisenberg model~\cite{bieri2016,Bieri-2015}. In a similar vein, it would be interesting to identify the QSL whose structure factor profile resembles that of the novel icosahedral order, and obtain a knowledge of its low-energy gauge structure, $U(1)$ or $\mathbb{Z}_{2}$, gapped vs gapless, etc. It would also be worthwhile to study the propensity towards dimerized states~\cite{Iqbal-2016_nem,Hering-2022,Iqbal-2019,Iqbal-2023_pp}, since such $J_{3}$ couplings on the kagome lattice are known to induce valence bond crystals in the vicinity of chiral spin liquids~\cite{Gong-2015}.

From a materials perspective, a number of experimentally studied natural minerals~\cite{Fennell-2011,Kampf-2013,mills-2014,Iqbal_Spangolite} and synthetic crystals~\cite{Cave-2006,Aliev-2012,Haraguchi-2018,Haraguchi-2021} have come into the limelight. Subsequent theoretical analysis of the complex frustration mechanism at play in these compounds is in a nascent stage, both as regards the nature of magnetic interactions and the consequences of their interplay~\cite{Makuta-2021,Ghosh-2023}. It can be envisaged that, either in synthesis of polymorphs, or in naturally occuring minerals, a scenario similar to that in the kagome materials kapellasite~\cite{Colman-2008} and haydeite~\cite{Boldrin-2015} plays out, whereby the nonmagnetic ions (in these cases Zn and Mg, respectively), occupy the centers of the hexagons, thus spanning the pairs of magnetic ions connected by $J_{3}$. The synthesis of compounds with such superexchange paths is likely to trigger a sizeable $J_{3}$ interaction, and depending on its strength could induce magnetic fluctuations displaying profiles of noncoplanar orders, as have been observed in kapellasite which displays cuboc-2 type magnetic fluctuations~\cite{Fak-2012}.\\


\paragraph*{Data availability.--}
The numerical data shown in the figures is available on Zenodo~\cite{gembe_2024_10658313}.
\\

\acknowledgments
We thank J.~Richter and K. Penc for discussions and joint work on related projects. 
The research of Y.I., C.H., and S.T. was carried out, in part, at the Kavli Institute for Theoretical Physics in Santa Barbara during 
the ``A New Spin on Quantum Magnets" program in summer 2023, supported in part by grant NSF PHY-2309135 to the Kavli Institute for Theoretical Physics (KITP). 
The work of Y.~I., S.~T., and C.~H. was performed, in part, at the Aspen Center for Physics, which is supported by National Science Foundation Grant No.~PHY-2210452 and Y.~I.\, was also supported by a grant from the Simons Foundation (1161654, Troyer). Y.~I.\ acknowledges support from the ICTP through the Associates Programme, from the Simons Foundation through Grant No.~284558FY19, IIT Madras through the Institute of Eminence (IoE) program for establishing QuCenDiEM (Project No.~SP22231244CPETWOQCDHOC), and the International Centre for Theoretical Sciences (ICTS), Bengaluru, India during a visit for participating in the program ``Kagome off-scale (Code No. ICTS/kagoff2024). Y.~I.\ also acknowledges the use of the computing resources at HPCE, IIT Madras. L.G. thanks IIT Madras for funding a three-month stay through an IoE International Graduate Student Travel award, where this project was initiated and worked on in the early stages.
M.G. thanks the Bonn-Cologne Graduate School of Physics and Astronomy (BCGS) for support.
The Cologne group acknowledges partial funding from the DFG within Project ID No. 277146847, SFB 1238 (projects C02, C03).
The numerical simulations were performed on the JUWELS cluster at the Forschungszentrum J\"ulich 
and the Noctua2 cluster at the Paderborn Center for Parallel Computing (PC$^2$).

\bibliography{mll_noncoplanar}

\clearpage
\appendix

\section{Supplementary material on the classical Monte Carlo simulations}
\label{app:classical_phase_diagram}

For the analysis of the classical phase diagram and the different ground states, we use a combination of classical Monte Carlo simulations in conjunction with a semi-analytical method, which we briefly explain in this appendix.

\subsection{Monte Carlo simulations}
\label{sec:mc}
All Monte Carlo simulations are performed on finite lattices of $L \times L$ unit cells with periodic boundary conditions, that is, $N = 6L^2$ spins ($L = 12$ unless stated otherwise). 
Local updates are performed with the Metropolis-Hastings algorithm. To resolve the thermal selection of ground states by thermal order-by-disorder effects at very low temperatures, we employ a parallel tempering/replica exchange Monte Carlo scheme \cite{hukushima1996exchange, swendsen1986replica} with 192 logarithmically spaced temperature points between $T_{\text{min}} = 10^{-4}$ and $T_{\text{max}} = 10$. These replicas are simulated simultaneously such that after every sweep, spin configurations of neighboring replicas are attempted to be exchanged according to some probability. As a result, the individual replicas perform a random walk in temperature space and can thus easily escape from local minima at low temperatures. We have taken care to check the thermalization of our parallel tempering scheme against feedback-optimized temperatures. \cite{katzgraber2006feedback}. For the specific heat data, a conventional Monte Carlo scheme without parallel tempering and 192 linearly spaced temperature points between $T_{\text{min}} = 0.02$ and $T_{\text{max}} = 0.5$ is employed. In all cases, measurements are performed over $5 \cdot 10^8$ sweeps after a thermalization period of $10^8$ sweeps. The static spin structure factors shown in the classical phase diagram (Fig.~\ref{Fig:classical_phase_diagram}) are obtained from the Fourier transform of the Monte Carlo equal-time real space spin-spin correlations, that is,
\begin{equation}
    S(\mathbf{q}) = \frac{1}{N} \sum_{i,j = 1}^N \langle \mathbf{S}_i \cdot \mathbf{S}_j\rangle e^{i\mathbf{q}\cdot(\mathbf{r}_i-\mathbf{r}_j)} \,,
\end{equation}
where $\mathbf{q}$ is a momentum inside the extended Brillouin zone, and $\mathbf{r}_i$ denotes the position of site $i$.

\begin{figure}[b!]
	\centering
	\includegraphics[width=1\columnwidth]{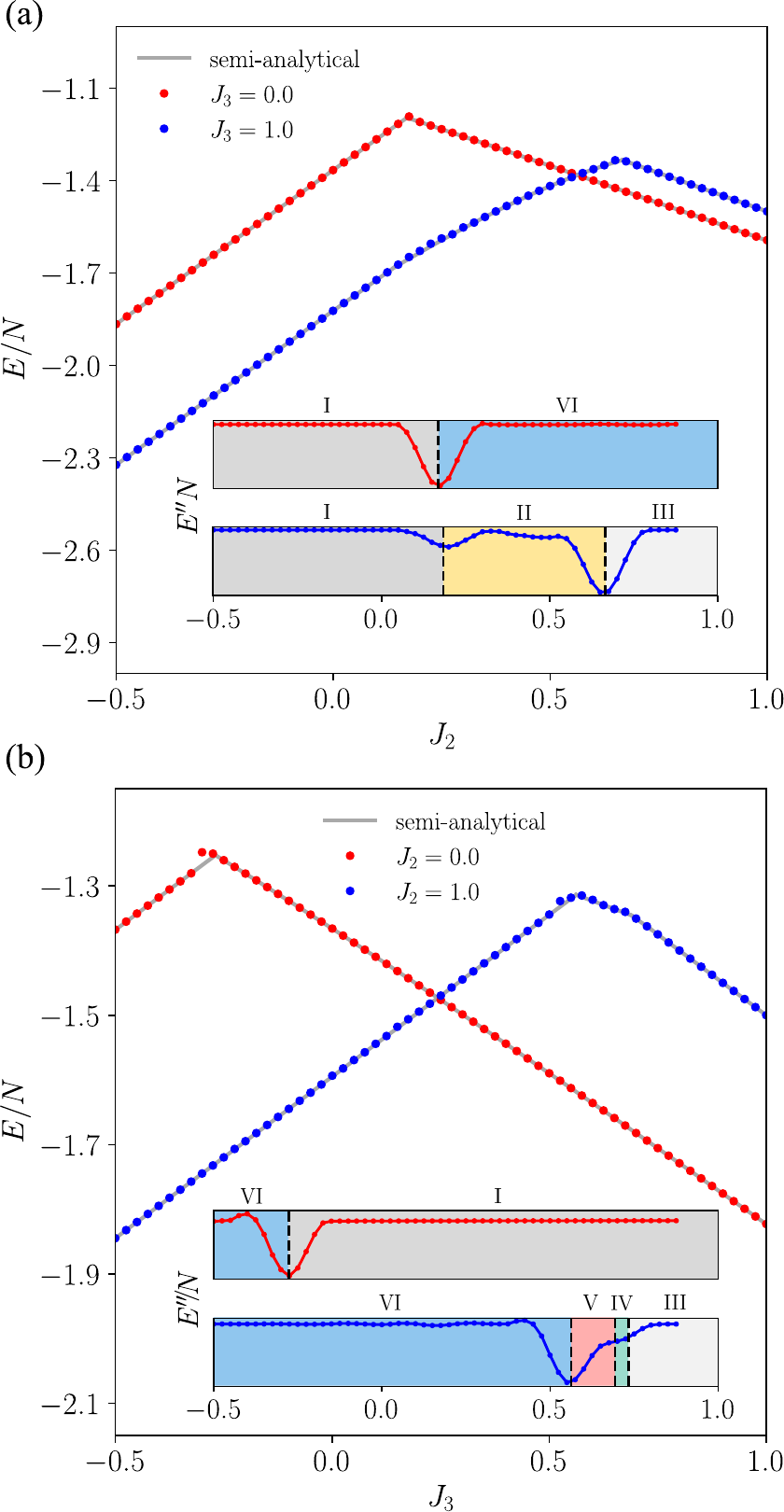}
	\caption{\textbf{Ground-state energies} from Monte Carlo simulations at low temperatures $(T = 10^{-4})$ for (a) fixed $J_3 = 0.0, 1.0$, and (b) fixed $J_2 = 0.0, 1.0$, respectively, compared to the semi-analytically obtained ground-state energies. The insets show the corresponding second derivatives, $E''/N$. These show clearly visible features exactly at the semi-analytically determined phase boundaries (indicated as vertical dashed lines).
	}
	\label{Fig:mll_energy_cuts}
\end{figure}

\subsection{Semi-analytical method}
\label{sec:sam}
We also use a semi-analytical approach presented in \cite{Gembe2023}, which works as follows: Starting from a ground state for a system of $N$ classical spins $\mathbf{S}_i$, which we obtain numerically from a Monte Carlo simulation, we first transform onto the eigenbasis of the corresponding tensor of inertia. Then we form groups of spin vectors that point approximately in the same direction, i.e. that fulfill, e.~g., $\mathbf{S}_i\cdot\mathbf{S}_j \ge 0.995$.
This results in a set of $M$ different spin directions, which is further reduced by guessing their symmetry group. In the end, we have $K$ different spin directions for the ground state, from which we obtain all others by applying symmetry operations. Next, we calculate the energy of the spin configuration
${\mathcal H}(\alpha_1,\ldots,\alpha_n)$ as a function of some parameters $\alpha_1,\ldots,\alpha_n$ that describe the position of the remaining $K$ spin vectors. These parameters have different meanings for the different phases, e.~g., they could be some $z-$ values that are constant for groups of spin vectors or certain difference angles between configurations of groups of spins. We need at most $n=2K-1$ parameters to characterize the spin configuration, but often less, see Table \ref{Tab:Summary}. The energy ${\mathcal H}(\alpha_1,\ldots,\alpha_n)$ is then numerically minimized starting with the initial numerical values of the parameters corresponding to the $K$ spin vectors.
It should have become clear that the semi-analytical method cannot be applied schematically, but requires a certain amount of intuition.
Incorrect identifications of closely neighboring spins are usually reflected in ground-state energies that are too high. Conversely, a slight lowering of the numerical initial energy is an indication of a successful application of the method.

\subsection{Supplemental numerical data}

Let us supplement the results for the classical phase diagram discussed in the main text with cuts of the Monte Carlo ground-state energy as a function of $J_2$ and $J_3$, which are presented in Fig.~\ref{Fig:mll_energy_cuts}. 

These underline the good interplay of Monte Carlo numerics on the one side, and semi-analytical method on the other side:
The insets in Fig.~\ref{Fig:mll_energy_cuts} display the second derivatives of the Monte Carlo ground-state energy where the
peaks indicate phase transitions along with the semi-analytically determined phase boundaries. 
Note that it is precisely at these boundaries that the Monte Carlo energies have clearly visible features, 
which is a confirmation of the accuracy of the (semi-analytically) determined phase boundaries. There is one slight deviation in the transition from phase VI to I, where the Monte Carlo energy shows a jump but the semi-analytic result stays continuous. This is likely due to insufficient thermalization to the very low temperature of $T = 10^{-4}$ in the presence of competing orders. This does, however, not affect any of the presented results in the main text. 

\section{Supplementary material on the Pseudo-Fermion Functional Renormalization Group}
\label{app:pf-FRG}

To calculate the quantum phase diagram in Fig.~\ref{Fig:quantum-phase-diagram} we employ pseudo-fermion functional renormalization group (pf-FRG) calculations. In this appendix, we shortly state the idea of the pf-FRG approach, and give references for readers interested in more details. We then describe our precise criterion for distinguishing disordered from ordered ground states used in analyzing the pf-FRG flow. Finally, we provide additional cuts through the quantum phase diagram for a better illustration of the transitions between the observed phases.

\subsection{Method}
The core concept of the pf-FRG involves avoiding the simultaneous treatment of all energy scales in the quantum problem at once. Instead, the approach starts at a known high-energy limit and subsequently incorporates lower energy scales in an iterative manner. To this end, an infrared cutoff, or RG Scale, $\Lambda$ is inserted into the model, so that in the high-energy limit $(\Lambda \to \infty)$ all correlation functions are completely determined by the bare couplings in the Hamiltonian, and in the low-energy limit $(\Lambda = 0)$ the full theory is recovered. In our case, the cutoff is implemented in Matsubara frequency space by multiplying the bare propagator with a continuous  regulator function. The interpolation between high and low energies is governed by an infinite hierarchy of differential equations, called \emph{flow equations}. Employing the Katanin truncation \cite{katanin2004}, we approximate this infinite hierarchy by a finite number of flow equations for the two- and four-point correlations, which we can---under certain approximations---solve numerically using the the \texttt{PFFRGSolver.jl} Julia package \cite{PFFRGSolver}. From the flow of these correlations we can then determine if the ground state of a given spin model is likely ordered or disordered. Details on this are given in the appendix below. For readers interested in more details on the pf-FRG approach we refer to the review \cite{muller2023}, and for more details on our specific implementation, we refer to \cite{kiese2022}.

\begin{figure}[t]
    \centering
    \includegraphics{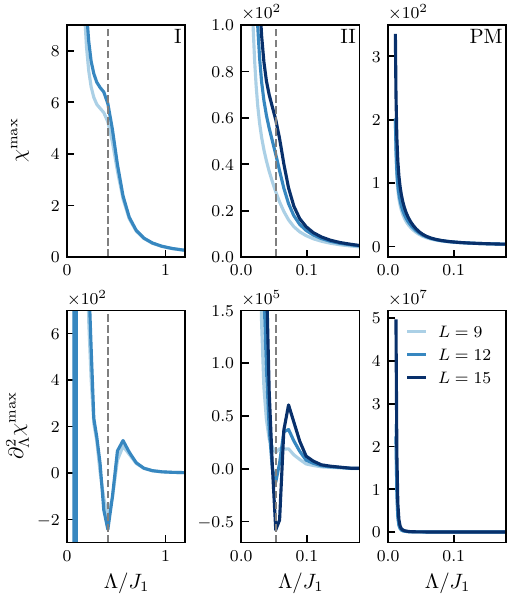}
    \caption{{\bf Renormalization group flow of the structure factor} in ordered (I and II) and disordered (PM) phases. The top row shows the structure factor flow at the momentum where it is maximal. The bottom row shows the second derivative of the flow with respect to the cutoff $\Lambda$, in which non-monotonic features signal the development of a flow breakdown. The position of the flow breakdown, the critical scale $\Lambda_c$, is depicted by the dashed gray lines. In the quantum disordered regime (right column), no flow breakdown occurs.}
    \label{fig:frg-flows}
\end{figure}

\subsection{Flow breakdown criterion}

To probe for magnetic order in the ground state of a given spin Hamiltonian, we  calculate the (static) spin-spin correlation $\chi_{ij}^\Lambda$ defined in Eq.~\eqref{eq:correlations}
from the two- and four-point correlations. We then Fourier transform $\chi_{ij}^\Lambda$ to obtain the flow of the (static) structure factor. If the ground state of the Hamiltonian under consideration exhibits magnetic order, this flow will show a divergence, or \emph{flow breakdown}, at a finite critical scale $\Lambda_c$ for the Bragg momenta characterizing the incipient order. If, on the other hand, there is no flow breakdown, the ground state is expected to be a disordered, putative quantum spin liquid state. In practice, the lattice truncation $L$ and the truncation of the flow equations will usually soften the divergence indicating a flow breakdown to a cusp or a peak, which becomes more prominent with increasing $L$. The flow of a disordered state, on the other hand, is expected to stay smooth and convex down to the lowest considered RG scale (in our case $\Lambda_\mathrm{min} = 0.01 |J|$, with $|J|^2 = J_1^2 + J_2^2 + J_3^2$ as normalization). We, therefore, identify any non-monotonicity in the second derivative of the structure factor flow as a flow breakdown, under the condition that it becomes more pronounced with increasing lattice size. For this comparison, we use up to three different truncation lengths $L = 9, 12, 15$. Examples of structure factor flows and their second derivative are shown in Fig~\ref{fig:frg-flows}. In the ordered regime, the second derivative shows a clear non-monotonicity, resulting in a cusp in the structure factor flow, and a clear lattice size dependence. In the disordered case, both the flow and its second derivative are smooth, monotonous, and essentially lattice size independent, signaling the absence of long-range order.

\subsection{Cuts through the quantum phase diagram}

\begin{figure*}
    \centering
    \includegraphics{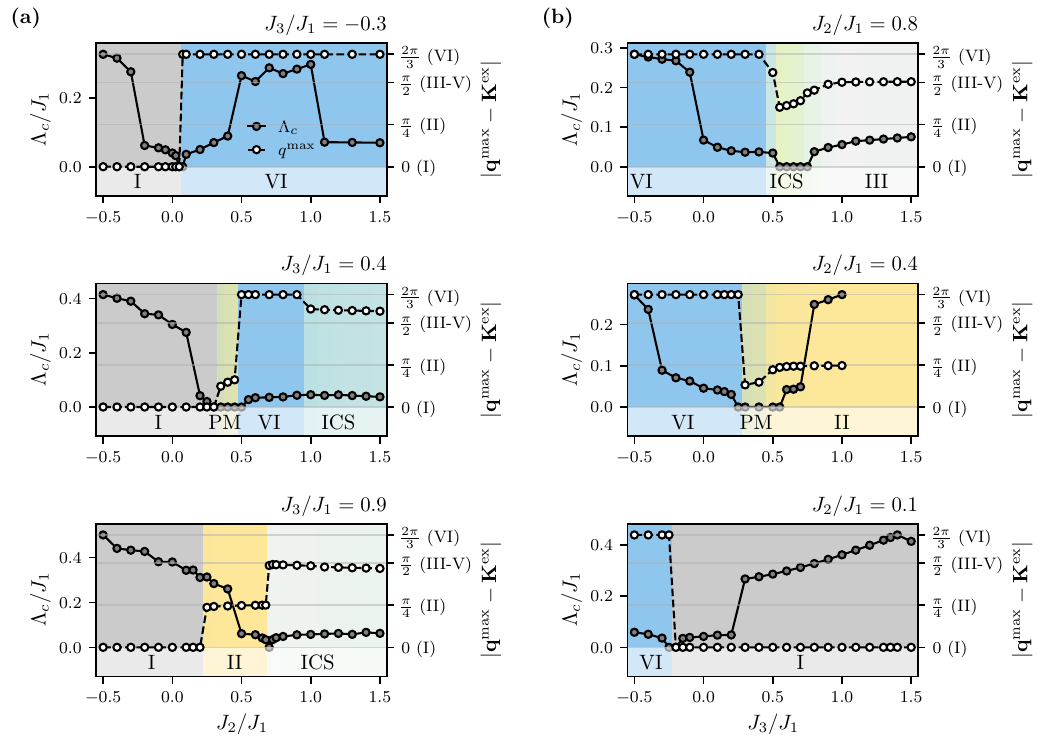}
    \caption{{\bf Cuts through the quantum phase diagram} shown in Fig.~\ref{Fig:quantum-phase-diagram}. Gray circles depict the critical scale $\Lambda_c$ and white circles the evolution of the ordering vectors $\mathbf{q}^\mathrm{max}$ for \textbf{(a)} vertical cuts ($J_3/J_1 = const.$) and \textbf{(b)} horizontal cuts ($J_2/J_1 = const.$). In addition to the ordered phases identified in our classical analysis, we observe large regions with ordering vectors at incommensurate (ICS) momenta.}
    \label{fig:quantum-phase-diagram-cuts}
\end{figure*}

For better interpretation of the full quantum phase diagram shown in  Fig.~\ref{Fig:quantum-phase-diagram}, we depict the evolution of the critical scale $\Lambda_c$ and the ordering vector $\mathbf{q}^\mathrm{max}$ along vertical (horizontal) cuts through parameter space with fixed  $J_3/J_1$ ($J_2/J_1)$ in Fig.~\ref{fig:quantum-phase-diagram-cuts}. 

This better illustrates the regions in parameter space with incommensurate order (ICS), where the ordering vector neither lies on a symmetry point of the first nor the extended Brillouin zone of the maple-leaf lattice. We also clearly see the continuous evolution of the ordering vector between phase III and VI, instead of the two distinct phases IV and V observed in the classical analysis (as visible in the upper left panel for fixed $J_2/J_1 = 0.8$). 

Additionally, we observe dips in the critical scale at the phase boundaries between phases I and VI, and phases II and III (or the nearby ICS phase), indicating a  phase transitions. Interestingly, the critical scale shows no notable feature at the boundary between phases I and II, suggesting a crossover instead of a phase transition. This would be contradictory to the different symmetries of the corresponding ground states identified in the classical analysis ($\mathrm{D}_3$ vs. $\mathrm{T}\times\mathds{Z}_2$). However, a similar cut showing the classical ground-state energy in Fig.~\ref{Fig:mll_energy_cuts} also shows only a very soft kink, indicating a very weak first-order transition that might not be captured well by just considering the critical scale of the pf-FRG. The pf-FRG ordering vectors, on the other hand, do show a sharp jump at the phase boundary, although this boundary is slightly shifted compared to the classical phase diagram.

\section{Unconstrained Luttinger-Tisza}
\label{app:unconstrained-lt}

\begin{figure*}
    \includegraphics{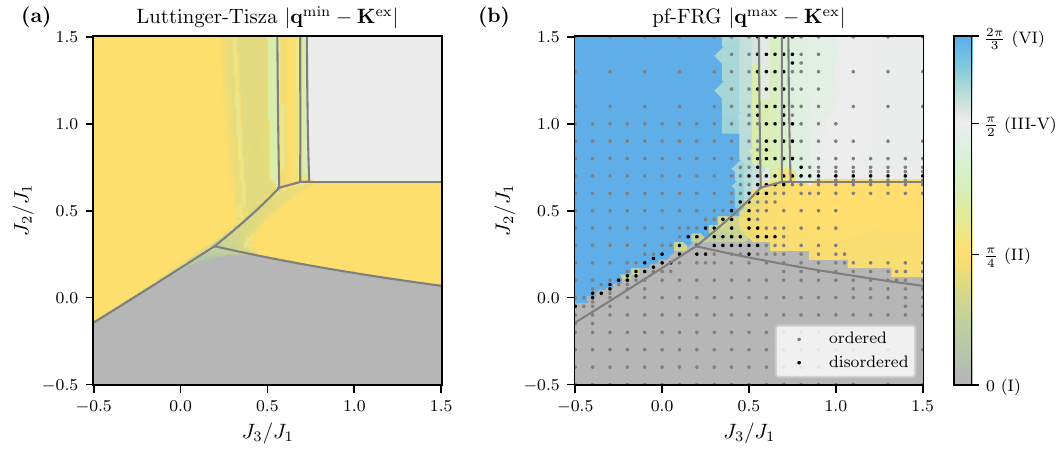}
    \includegraphics{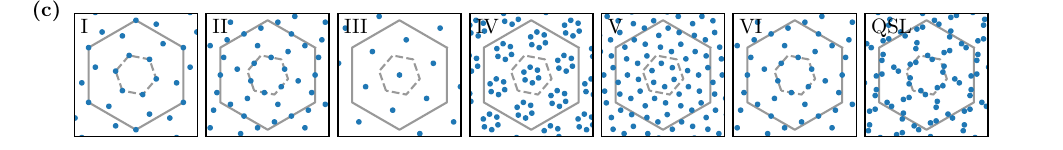}
    \caption{{\bf Unconstrained Luttinger-Tisza versus pf-FRG}. \textbf{(a)} LT $\mathbf{q}$-vectors with minimal eigenvalue of the Fourier transformed interaction matrix. \textbf{(b)} Distance of the momentum where the pf-FRG structure factor is maximal $\mathbf{q}^{max}$ to the $\mathbf{K}^\mathrm{ex}$-point of the extended Brillouin zone. Quantum disordered regions that show no flow breakdown $(\Lambda_c = 0)$ are marked by black dots. The gray lines are the classical phase boundaries. \textbf{(c)} All $\mathbf{q}$-vectors with minimal Luttinger-Tisza eigenvalue at the same parameters as shown for the quantum case (Fig.~\ref{Fig:quantum-phase-diagram}). The dashed and solid gray lines show the first and  extended Brillouin zone, respectively. Note that for the $\mathbf{q}$-vectors depicted in phase IV and V the $\emph{hard}$ spin length constraint is not fulfilled, and thus they don't necessarily describe the true classical ground state. For the momenta depicted in all other phases the constraint is fulfilled and characterizes the exact classical ground state.}
    \label{Fig:lt-phase-diagram}
\end{figure*}

\begin{figure*}
    \centering
    \includegraphics{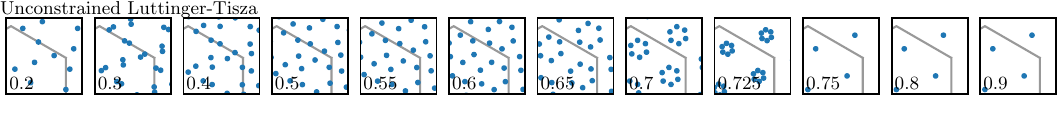}
    \includegraphics{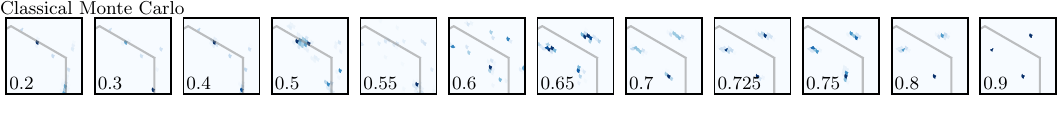}
    \caption{{\bf Ground-state $\mathbf{q}^\mathrm{min}$-vectors from unconstrained Luttinger-Tisza and structure factors from Monte Carlo} evolving from phase VI to III for fixed $J_2/J_1 = 1$ and increasing $J_3/J_1$ (indicated in the bottom left corners). Similar to the pf-FRG results (c.f. Fig.~\ref{fig:quantum-sfs-transition}), the LT analysis suggests a seemingly continuous evolution with no clearly distinct phases IV and V that were identified in our classical analysis. The classical Monte Carlo calculation does not properly capture this evolution due to the incommensurate nature of the ground state being incompatible with a finite periodic lattice.
    Note, however, that the \emph{hard} spin length constrained is not fulfilled in these incommensurate phases for any of the depicted LT $\mathbf{q}$-vectors, and thus they don't necessarily describe the true classical ground state.} \label{fig:lt-qvectors-transition}
\end{figure*}

In order to substantiate the structure factors derived from our pf-FRG calculations, especially within the incommensurate regimes where they disagree with the classical analysis, we employ the unconstrained Luttinger-Tisza (LT) approach \cite{luttinger1946, luttinger1951}. This approach studies the classical model, but relaxes the hard constraint of constant spin length on each spin to a constraint on the total spin length. Enforcing only this \emph{weak constraint} enables a straight-forward Fourier transformation of the interaction matrix, and a subsequent diagonalization of the Hamiltonian in momentum space. The momenta $\mathbf{q}^\mathrm{min}$ with eigenvectors of minimal energy then characterize the LT ground state---a semi-classical approximation to the true ground state with an energy that serves as a lower bound on the exact ground-state energy. The fact that the LT approach works on an infinite lattice and that, as in the quantum model, spin length is not conserved, makes it a suitable tool to compare it to and corroborate the results from our pf-FRG calculation. 

Figure~\ref{Fig:lt-phase-diagram} does just that, depicting the LT $\mathbf{q}$-vector with minimal eigenvalue $\mathbf{q}^\mathrm{min}$ \textbf{(a)} next to the pf-FRG ordering vectors $\mathbf{q}^\mathrm{max}$ of maximal structure factor intensity \textbf{(b)}, and shows examples of the full LT ground-state $\mathbf{q}$-vectors \textbf{(c)} in the different phases. The most notable difference is that the LT $\mathbf{q}$-vectors are periodic with the reciprocal lattice vectors of the triangular Bravais lattice of the maple-leaf lattice, and thus fully specified by points in the first Brillouin zone (depicted by the dashed lines in Fig.~\ref{Fig:lt-phase-diagram}(c)). 
Conversely, the spin structure factor as calculated from the pf-FRG and Monte Carlo are periodic with the reciprocal lattice of the smaller triangular lattice that, when depleted by 1/7, transforms into the maple-leaf lattice. They are thus fully specified by points in the extended Brillouin zone (solid gray lines). This explains the resulting additional $\mathbf{q}$-vectors found by the LT approach not present in the spin structure factor. Due to these extra momenta, phase II and phase VI appear equivalent in the unconstrained LT approach, whereas both the structure factor from pf-FRG and the classical analysis only show a subset the $\mathbf{q}$-vectors and reveal that the phases indeed differ even in their respective ground-state symmetry (T vs $\mathrm{D}_3$).

Comparing only the LT $\mathbf{q}$-vectors also present in the pf-FRG structure factor, however, the two methods agree remarkably well, both showing incommensurate $\mathbf{q}$-vectors in the paramagnetic regime and in between phases III and VI. Our LT calculation also predicts the smooth evolution of the structure factor from phase III to VI, as depicted in Fig.~\ref{fig:lt-qvectors-transition}, resembling the pf-FRG structure factors in Fig.~\ref{fig:quantum-sfs-transition}. This supports the incommensurate nature of the ground state in this parameter regime, which was not seen by our initial classical analysis. For comparison, Fig.~\ref{fig:lt-qvectors-transition} also depicts structure factors from classical Monte carlo, which don't properly capture the continuous evolution.

To compare the LT approach with our classical analysis, we have also checked for which of the LT $\mathbf{q}$-vectors the \emph{hard} spin constraint is fulfilled. We find that this is true for all phases expect phases IV and V. The $\mathbf{q}$-vectors in all other phases therefore characterize the exact classical ground state. To fulfill the constraint in the noncoplanar phases II and VI, three different momenta of the ones depicted in Fig.~\ref{Fig:lt-phase-diagram}(c) have to be chosen for each spin dimension, e.g. $(0, \pi)$ for $S^x$, $(\pi, 0)$ for $S^y$ and $(\pi, \pi)$ for $S^z$, as also suggested in TABLE.~\ref{Tab:Summary} (where the momenta are given in the basis for which the reciprocal lattice vectors are $(2\pi, 0)$ and $(0, 2\pi))$. This makes it possible to construct a noncoplanar real-space spin configuration that fulfills the hard spin constraint.
\clearpage

\begin{figure*}
    \centering
    \includegraphics{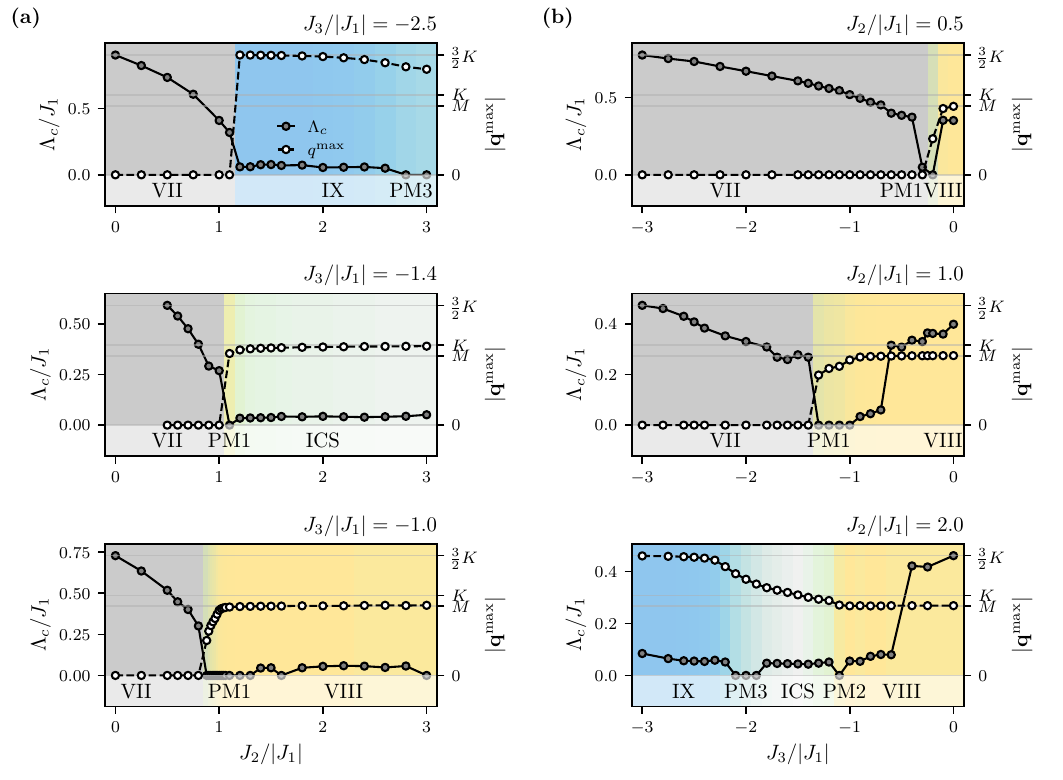}
    \caption{{\bf Cuts through the quantum phase diagram for ferromagnetic $J_1 < 0$} shown in Fig.~\ref{Fig:quantum-phase-diagram-fm}. Gray circles depict the critical scale $\Lambda_c$ and white circles the evolution of the ordering vectors $\mathbf{q}^\mathrm{max}$ for \textbf{(a)} vertical cuts ($J_3/J_1 = const.$) and \textbf{(b)} horizontal cuts ($J_2/J_1 = const.$).}
    \label{fig:quantum-phase-diagram-cuts-fm}
\end{figure*}
\begin{figure*}
    \centering
    \includegraphics{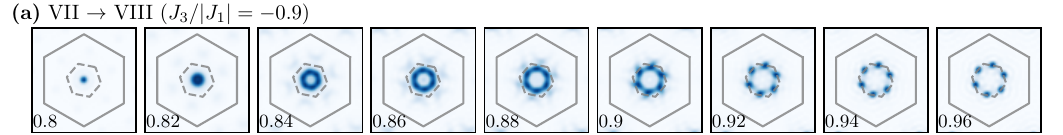}
    
    \vspace{0.2cm}
    
    \includegraphics{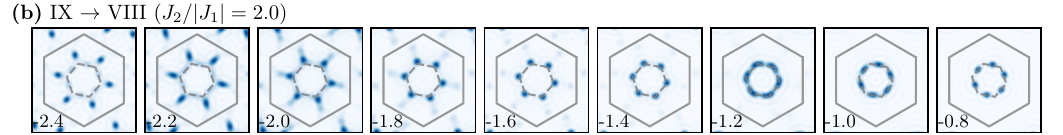}
    
    \caption{{\bf Ground-state structure factors from pf-FRG in the incommensurate regime} for ferromagnetic $J_1 < 0$ evolving \textbf{(a)} from phase VII to VIII for fixed $J_3/|J_1| = -0.9$ and increasing $J_3/|J_1|$ (indicated in the bottom left corners) and \textbf{(b)} from phase XI to VIII for fixed $J_2/|J_1| = 2.0$ and increasing $J_3/|J_1|$. In both cases the structure factor evolves continuously from one phase into the other, showing peaks at incommensurate momenta in between.}
    \label{fig:quantum-sfs-fm-transitions}
\end{figure*}
%

\section{Supplemental material for the ferromagnet}
\label{app:fm}

In this appendix we provide additional data for the FRG calculation as well as a phase diagram from unconstrained Luttinger Tisza to substantiate our discussion regarding the classical and quantum phase diagram of the ferromagnet in section~\ref{sec:ferromagnet}.

\subsection{Quantum structure factors and cuts through the quantum phase diagram}

For better interpretation of the full quantum phase diagram shown in  Fig.~\ref{Fig:quantum-phase-diagram-fm}, Fig.~\ref{fig:quantum-phase-diagram-cuts-fm} depicts the evolution of the critical scale $\Lambda_c$ and the ordering vector $\mathbf{q}^\mathrm{max}$ along vertical (horizontal) cuts through parameter space with fixed  $J_3/J_1$ ($J_2/J_1)$. Most notably, these reveal the continuous evolution of the structure factor in the incommensurate regimes. The lower right panel also illustrates that the incommensurate region between phases PM2 and PM3, only exhibits structure factor peaks exactly at the commensurate $\mathbf{K}$-vector at a single point along the cut in parameter space, and no extended commensurate region is found. The continuous evolution of the structure factor in the incommensurate regimes between phases VII and VIII, and phases VIII and IX, are shown in Fig.~\ref{fig:quantum-sfs-fm-transitions}.

\begin{figure*}
    \includegraphics{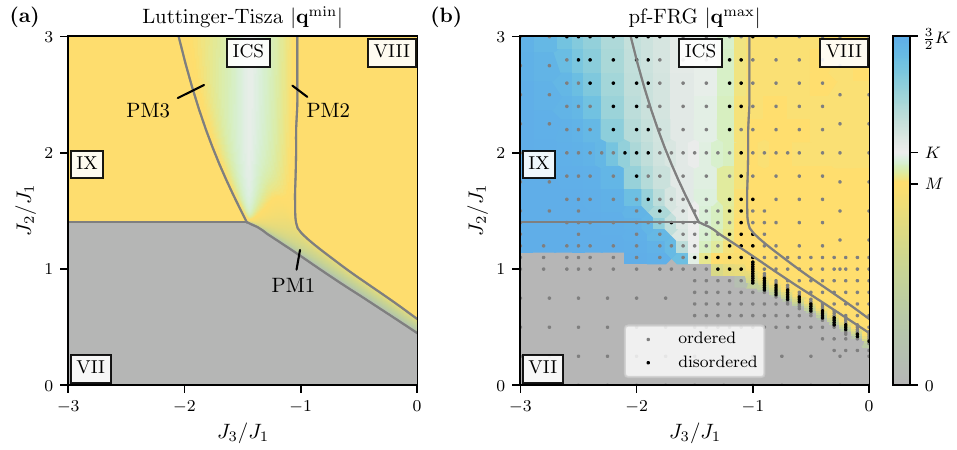}
    \includegraphics{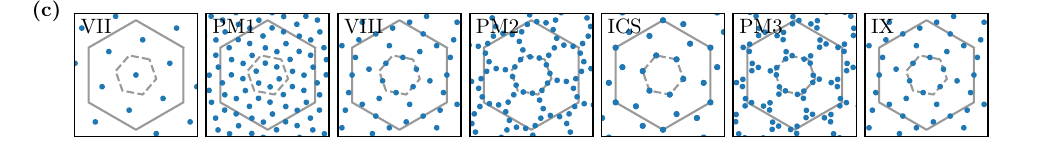}
    \caption{{\bf Unconstrained Luttinger-Tisza versus pf-FRG for ferromagnetic $J_1 < 0$}. \textbf{(a)} LT $\mathbf{q}$-vectors with minimal eigenvalue of the Fourier transformed interaction matrix. \textbf{(b)} Distance of the momentum where the pf-FRG structure factor is maximal $\mathbf{q}^{max}$ to the $\Gamma$-point of the  Brillouin zone. Quantum disordered regions that show no flow breakdown $(\Lambda_c = 0)$ are marked by black dots. The gray lines are the classical phase boundaries. \textbf{(c)} All $\mathbf{q}$-vectors with minimal Luttinger-Tisza eigenvalue deep into the different phases VII, VIII and IX, and in the ICS regime separating them. The annotations and labels indicate the parameters for which the $\mathbf{q}$-values are calculated. The dashed and solid gray lines show the first and  extended Brillouin zone, respectively. Note that for the $\mathbf{q}$-vectors in the ICS regime the spin length constraint is not fulfilled, and thus they don't necessarily describe the true classical ground state. For the momenta depicted in all other phases the constraint is fulfilled and characterizes the exact classical ground state.}
    \label{Fig:lt-phase-diagram-fm}
\end{figure*}
\begin{figure*}
    \centering
    \includegraphics{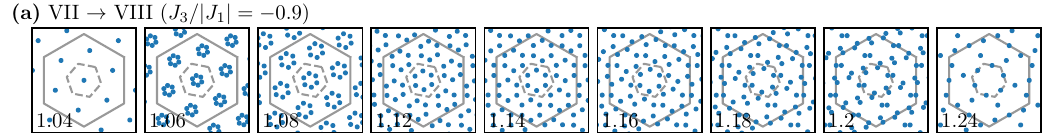}
    
    \vspace{0.1cm}
    
    \includegraphics{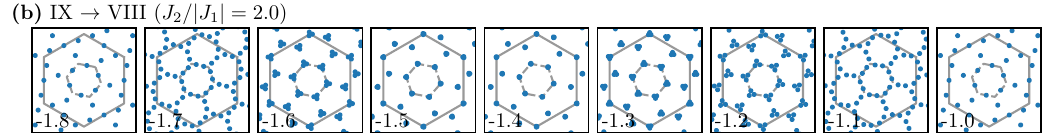}
    
    \caption{{\bf Ground-state $\mathbf{q}^\mathrm{min}$-vectors from unconstrained Luttinger-Tisza for ferromagnetic $J_1 < 0$} evolving \textbf{(a)} from phase VII to VIII for fixed $J_3/|J_1| = -0.9$ and increasing $J_3/|J_1|$ (indicated in the bottom left corners) and \textbf{(b)} from phase XI to VIII for fixed $J_2/|J_1| = 2.0$ and increasing $J_3/|J_1|$. Similar to the pf-FRG results (c.f. Fig.~\ref{fig:quantum-sfs-fm-transitions}), the LT analysis suggests a seemingly continuous evolution of the structure factor from one phase to the other through a regime showing incommensurate order. 
    Note, however, that the \emph{hard} spin length constrained is not fulfilled in these incommensurate phases for any of the depicted LT $\mathbf{q}$-vectors, and thus they don't necessarily describe the true classical ground state.}
    \label{fig:lt-qvectors-fm-transitions}
\end{figure*}
\subsection{Unconstrained Luttinger Tisza}
\label{sec:unconstrained-lt-fm}
As for the antiferromagnet, we additionally calculate a phase diagram from unconstrained Luttinger-Tisza, which also shows the evolution of the $\mathbf{q}$-vectors in the incommensurate regime were the hard spin constraint is not fulfilled. Fig.~\ref{Fig:lt-phase-diagram-fm} shows the absolute value of the resulting $\mathbf{q}$-vectors in comparison to the pf-FRG structure factor peaks, as well as all $\mathbf{q}$-vectors in the different observed phases.  Fig~\ref{fig:lt-qvectors-fm-transitions} shows the evolution of the $\mathbf{q}$-vectors from phase VII to VIII and IX to VIII, in analogy to the quantum version depicted in Fig.~\ref{fig:quantum-sfs-fm-transitions}. We again see a good agreement with the pf-FRG structure factors, apart from the different periodicity (as discussed in the previous sections), and a notable shift of the phase boundaries.


\end{document}